\documentclass[preprint,authoryear,12pt]{elsarticle}
\usepackage{verbatim}  % Needed for the "comment" environment 
\usepackage{graphicx}
\usepackage{color}
\usepackage{hyperref}
\graphicspath{{Figures/}}  % Location of the graphics files (set up for graphics
\usepackage{amssymb}
\usepackage{multirow}
\usepackage{rotating}
\usepackage{color}
\usepackage{booktabs}
\usepackage{threeparttable}

\journal{Computers, Environment and Urban Systems}

\begin{document}

\begin{frontmatter}% Title, authors and addresses

\title{\tnoteref{title}}

\author{Robin Lovelace}
\ead{robin.lovelace@shef.ac.uk}

\author{Dimitris Ballas}

\address{Department of Geography,
The University of Sheffield,
Sheffield,  S10 2TN,
United Kingdom}



\title{`Truncate, replicate, sample': a method for creating integer weights for
spatial microsimulation}

\begin{abstract}
Iterative proportional fitting (IPF) is a widely
used method for spatial microsimulation.
The technique results in non-integer weights for individual rows of
data. This is problematic for certain applications and has led many researchers
to favour combinatorial optimisation approaches such as simulated annealing. An
alternative to this is `integerisation' of IPF weights: the translation of the
continuous weight variable into a discrete number of unique or `cloned'
individuals. We describe four existing methods of integerisation and
present a new one. Our method --- `truncate, replicate, sample' (TRS) ---
recognises  that IPF weights consist of both
`replication weights' and `conventional weights', the effects of which need to
be separated. The procedure consists of three steps: 1)
separate replication and conventional weights by
truncation; 2) replication of individuals with
positive integer weights; and 3) probabilistic sampling.
The results, which are reproducible using supplementary
code and data published alongside this paper, show that TRS is fast, and more
accurate than alternative  approaches  to integerisation.
\end{abstract}
\end{frontmatter}

%%\tableofcontents
%%\listoffigures

% % ``It is humbly suggested that it is about time that quantitative geographers
% % started to devise a body of relevant spatial analysis techniques
% % that can cope with geographical data. The first stage in this second
% % quantitative revolution must be the development of methods that can handle the
% % MAUP'' \citep{Openshaw1983}

\section{Introduction}
%Why do microsimulation?
% History
%Benefits of IPF (a little on what it is)
% Spatial variation is ubiquitous. A wide range of physical and
% social phenomena, from neuron connections in the brain to the structure of
% cities, display general spatial features. Many such spatial patterns can be
% analysed using a shared set of concepts and mathematical models
% \citep{Marc2011}. Similarly, the characteristics of points and polygons
% scattered over a two dimensional plane (whether house prices, soil moisture
% content or migration patterns between cities) tend to vary from place to place,
% and display spatial patterns \citep{Fik2003,Western1999,Rae2009}.
% This ubiquity has led to discussion about how to deal with space in a number of
%  disciplines, including Geography \citep{Sack1973}, Ecology \citep{Tilman1997},
% and Economics \citep{Rey2005}.
% 
% As these discussions have developed, the variety of methods available for the
% analysis of spatial data has grown. Researchers can now choose between
% techniques ranging simple chloropleth maps of zonal aggregates to complex
% space-time models (refs).

Spatial microsimulation has been widely and increasingly used as a term
to describe a  a set of techniques used to estimate the
characteristics of individuals within geographic zones about which only
aggregate statistics are available \citep{Tanton2012, Ballas2013}.
The model inputs operate on
a different level from those of the outputs. To ensure that the individual-level
output matches the aggregate inputs, spatial microsimulation mostly relies on
one of two methods. \emph{Combinatorial optimisation} algorithms are used to
select a unique combination of individuals from a survey dataset. This approach
was first demonstrated and applied by \citet{Williamson1998} and there have
been several applications and refinements since then. Alternatively,
\emph{deterministic reweighting} %Add tramner ref here! 
iteratively alters an array of weights, $N$, for
which columns and rows correspond to zones and individuals, to optimise the fit
between observed and simulated results at the aggregate level. This approach
has been implemented using iterative proportional fitting (IPF) to combine
national survey data with small area statistics tables
(e.g.~\citealp{Beckman1996, Ballas2005c}). A recent review, published in this
journal, highlights
the advances made in methods for simulating spatial microdata
\citep{Hermes2012a} since these works were published. \citet{harland2012} also
discuss the state of spatial microsimulation research
and present a comparative critique of the performance
of deterministic reweighting and combinatorial optimisation methods.
Both approaches require micro-level and spatially aggregated input data and
a predefined exit point: the fit between simulated
and observed results improves, at a diminishing rate, with each
iteration.\footnote{In
IPF, model fit improves from one iteration
to the next. Due to the selection of random individuals in simulated
annealing, the fit can get worse from one iteration to the next
\citep{Williamson1998, Hynes2009}. It is impossible to predict the final model fit in both
cases. Therefore exit points may be somewhat arbitrary. For IPF, 20 iterations
has been used as an exit point \citep{Lee2009, Anderson2007}. For simulated
annealing, 5000 iterations have be used \citep{Hynes2009, Goffe1994}.

\nopagebreak
}

The benefits of IPF include speed of computation, simplicity and the guarantee
of convergence \citep{Deming1940, Mosteller1968, Fienberg1970, Wong1992,
 Pritchard2012}. A
major potential disadvantage, however, is that  non-integer
weights are produced: fractions of individuals are present in a given
area whereas after combinatorial optimisation, they are either present or absent.
Although this is not a problem for many static
spatial microsimulation applications (e.g.~estimating income
at the small area level, at one point in time; for example
see \citealt{Anderson2013}), several applications require
integer rather than fractional
weights. For example, integer weights are required if a population is to be
simulated dynamically into the future (e.g. \citealp{Ballas2005c, Clarke1986,
Holm1996, Hooimeijer1996}) or linked to agent-based models
(e.g. \citealp{Birkin2011, Gilbert2008-abm, Gilbert2005, Wu2008,
Pritchard2012}).

Integerisation solves this problem by converting the weights ---
a 2D array of positive
real numbers ($N \in \mathbb{R}_{\geq0}$) --- into an array of
integer values \mbox{($N' \in \mathbb{N}$)} that represent whether
the associated individuals are present (and how many times they are replicated)
or absent.
The integerisation function
must perform $f(N) = N'$ whilst minimizing the difference between constraint
variables and the aggregated results of the simulated individuals.
Integerisation has been performed on the results of the SimBritain
model, based on simple rounding of the weights and two
deterministic algorithms that are evaluated subsequently in this paper
(see \citealp{Ballas2005c}). It was found that
integerisation ``resulted in an increase of the difference
between the `simulated' and actual cells of the target variables''
\citep[p.~26]{Ballas2005c}, but there was no further analysis of the amount of
error introduced, or which integerisation algorithm performed best.

To the best of our knowledge, no published research has quantitatively
compared the effectiveness of different integerisation strategies.
We present a new method --- truncate, replicate sample (TRS) --- that combines
probabilistic and deterministic sampling to generate representative integer
results. The performance of TRS is evaluated alongside four alternative methods.

  %The main purpose of
%this paper is to present an alternative method of integerisation which is fast,
%imple, and yields accurate results, compared with the method published by
%\citet{Ballas2005c}.
% The findings show that the TRS method for
% integerisation presented in this paper performs better than the alternatives,
% in terms of speed and accuracy.

% populate
% 
% 
% Despite this flourishing of options, many analyses
% suffer from the modifiable areal unit problem and
% the ecological fallacy, especially when the data are only provided as spatial
% aggregates to maintain confidentiality. Spatial microsimulation is a tool that
% can help overcome these problems by estimating the characteristics of
% individuals within areas about which only zonal aggregates are know. Iterative
% proportion fitting (IPF) is a simple, effective and frequently used technique
% for reweighting survey data at the individual level to fit geographical
% aggregates. However, IPF generates non-integer weights: an individual row of
% data may be counted 0.235 times, making the analysis problematic. This has
% led to attempts at integerisation, the conversion of non-integer weights
% into a unique combination of whole individuals, for each area under
% investigation.

An important feature of
this paper is the provision of code and data
that allow the results to be tested and replicated using the
statistical software R \citep{R2012}.\footnote{The code,
data and instructions to replicate the findings are
provided in the Supplementary
Information:
\url{https://dl.dropbox.com/u/15008199/ints-public.zip} . A larger open-source
code project, designed to test IPF and related algorithms under a range of
conditions, can be found on github:
\url{https://github.com/Robinlovelace/IPF-performance-testing} .
}
Reproducible research can be defined as that which allows others to
conduct at least part of the analysis (Table \ref{table:rep}).
Best practice
is well illustrated by
\citet{Williamson2007}, an instruction
manual on combinatorial optimisation algorithms
described in previous work. Reproducibility is
straightforward to achieve \citep{Gentleman2007}, has
a number of important benefits \citep{Ince2012}, yet is
often lacking in the field.

\begin{table}[h*]
\caption{Criteria for reproducible research, adapted from \citet{Peng2006}}
\begin{tabular}{p{4cm} p{9cm}}
\toprule
Research component & Criteria \\ \midrule
Data & Make dataset available, either in original form or in anonymous,
scrambled form if confidential \\ 
Methods & Make code available for data analysis. Use non-prohibitive software if
possible \\  
Documentation & Provide comments in code and describe how to replicate results
\\ 
Distribution & Provide a mechanism for others to access data, software, and
documentation \\ \bottomrule
\end{tabular}
\label{table:rep}
\end{table}

The next section reviews the wider context of spatial microsimulation
research and explains the importance of integerisation.
The need for new methods is established in Section
\ref{strategies}, which describes increasingly sophisticated
methods for integerising the results of IPF. Comparison of these
five integerisation methods show TRS to be more accurate
than the alternatives, across a range of 
measures (Section \ref{results}). The
implications of these findings
are discussed in Section \ref{discuss}.

\section{Spatial microsimulation: the state of the art}
\label{art}

\subsection{What is spatial microsimulation, and why use it?}
\label{smsim}
Spatial microsimulation is a modelling method that involves sampling rows of
survey data (one row per individual, household, or company) to generate lists of
individuals (or weights) for geographic zones that expand the survey to the
population of each geographic zone considered. The
problem that it overcomes is that most publicly available
census datasets are aggregated, whereas individual-level data are sometimes
needed. The ecological fallacy \citep{Openshaw1983}, for example, can be tackled
using individual-level data.

Microsimulation cannot replace the `gold standard' of real,
small area microdata \citep[p.~4]{Martin2002},
yet the method's practical usefulness (see
\citealp{Tomintz2008}) and testability \citep{Edwards2009} are beyond doubt.
With this caveat in mind, the
challenge can be reduced to that of optimising the fit between
the aggregated results of simulated
spatial microdata and aggregated census variables such as age
and sex \citep{Williamson1998}. These variables are often
referred to as `constraint variables' or `small area constraints'
\citep{Hermes2012a}. The term `linking variables' can also be used, as they
\emph{link} aggregate and survey data.

% Before undertaking spatial microsimulation, it is important to consider
% alternatives: Are aggregated data sufficient? Why not \emph{find out}
% peoples' characteristics in different areas using established sampling
% techniques \citep{Cochran1977}? First,
% spatial microsimulation takes full advantage of secondary data,
% avoiding expensive surveys \citep{Lee2009}. Second, the characteristics of
% \emph{all} individuals in each zone are simulated. This provides a
% basis for comprehensive
% agent-based models \citep{Ryan2009}.  Third,
% spatial microdata can provide estimates of
% `target variables' such as income at individual and local levels
% \citep{Ballas2005b}.\footnote{Regression
% estimator techniques, based on individual-level data, can also be used to
% estimate the value of target variables \citep{Cochran1977}. However, this
% technique relies on individual-level spatial microdata, which are not
% generally available for small zones (although they could be simulated by
% spatial microsimulation).
% }
% Other methods of researching intra-zone variability exist, such as calculating
% the level of inequality between geographic subdivisions \citep{Whitworth2011}.
% However, techniques based on aggregated data can mask variability at the
% individual level: high and low income households, for example, can live
% side-by-side \citep{Fahmy2008}.

The wide range of methods available for spatial microsimulation can be divided
into static, dynamic, deterministic and probabilistic approaches (Table
\ref{typology}). Static approaches generate small
area microdata for one point in time. These can be classified as
either probabilistic methods which use a random number generator, and
deterministic reweighting methods, which do not. The latter produce
fractional weights. Dynamic approaches project small
area microdata into the future. They typically involve modelling of
life events such as births, deaths and migration on the basis of random
sampling from known probabilities on such events \citep{Ballas2005c,
Vidyattama2010}; more advanced agent-based techniques, such as spatial
interaction models and household-level phenomena, can be added to this basic
framework \citep{Wu2008, Wu2010}. There
are also `implicitly dynamic' models, which employ a static
approach to reweight an existing microdata set to match
projected change in aggregate-level variables
(e.g.~\citealp{Ballas2005-ireland}).

\begin{table}[h]
\centerline{}
\caption{Typology of spatial microsimulation methods}
\vspace{0.25 cm}
\footnotesize{
\begin{tabular}{p{1.6cm}p{2.4cm}p{3.5cm}p{3.0cm}p{2cm}}
\toprule
{Type} & {Reweighting technique} & {Pros} & {Cons} &
{Example} \\ \midrule
\multirow{3}{2cm}{\vspace{0.3cm} \\ Determ-  inistic\\\vspace{0.3cm}
Re-\\weighting} & Iterative proportional fitting (IPF) & Simple, fast, accurate,
avoids local optima and random numbers & Non-integer weights &
\citep{Tomintz2008}⁠⁠ \\ \cmidrule{2- 5}
& Integerised IPF & Builds on IPF, provides integer
weights & Integerisation reduces model fit & \citep{Ballas2005c}⁠ \\
\cmidrule{2- 5}
&  GREGWT, generalised reweighting 
& Fast, accurate,
avoids local optima and random numbers & Non-integer weights  &
\citep{Miranti2010}⁠ \\ \midrule
\multirow{2}{2cm}{ \\ Probab- ilistic  Combin-
atorial optim-
isation} & Hill climbing approach & The simplest solution to a combinatorial
optimisation, integer results & Can get stuck in local optima, slow &
\citep{Williamson1998}⁠ \\ \cmidrule{2-5}
& Simulated annealing & Avoids local minima, widely
used, multi-level constraints & Computationally intensive
& \citep{kavroudakis2012}⁠  \\ \midrule
\multirow{2}{2cm}{\vspace{0.3cm} \\ Dynamic} & Monte Carlo
randomisation to simulate ageing  & Realistic treatement of stochastic
life events such as death & Depends on accurate estimates of life event
probabilities & \citep{Vidyattama2010}⁠ \\
\cmidrule{2- 5}
& Implicitly dynamic & Simplicity, low
computational demands & Crude, must project constraint
variables & \citep{Ballas2005b}⁠ \\ \bottomrule
\end{tabular}
}
\label{typology}
\end{table}

\subsection{IPF-based Monte Carlo approaches for the generation of synthetic
microdata}
Individual-level, anonymous samples from major surveys, such as the Sample of
Anonymised Records (SARs) from the UK Census have only been available since
around the turn of the century \citep{Li2004}. Beforehand, researchers had to
rely on synthetic  microdata. These can be created
using probabilistic methods \citep{Birkin1988}.
The iterative proportional fitting (IPF) technique was
first described in 1940 \citep{Deming1940}, and has become well established
for spatial microsimulation \citep{Muller2010, Birkin1989}.

The first application of IPF in spatial microsimulation was presented by Birkin
and Clarke (\citeyear{Birkin1988} and \citeyear{Birkin1989}) to generate 
synthetic individuals, and allocate them to small areas based on
aggregated data. They produced spatial microdata (a list of individuals and
households for each electoral ward in Leeds Metropolitan District). Their
approach was to select rows of synthetic data using Monte Carlo sampling.
Birkin and Clarke suggested that the microdata generation technique
known as `population synthesis' could be
of great practical use \citep{Birkin2012-geodem}.

\subsection{Combinatorial optimisation approaches}
Since the work of Birkin and Clarke (\citeyear{Birkin1988} and
\citeyear{Birkin1989}) there have been considerable advances in
data availability and computer hardware and software. In particular, with the
emergence of anonymous survey data, the focus of spatial
microsimulation shifted towards methods for reweighting and sampling from
existing microdata, as opposed to the creation of entirely synthetic data
\citep{Lee2009}.

This has enabled experimentation with new techniques for small area
microdata generation. A
significant contribution to the literature was made by \citet{Williamson1998}.
The authors presented
microsimulation as a problem of \emph{combinatorial optimisation}: finding the
combination of SARs which best fits the constraint variables.
Various approaches to combinatorial
optimisation were compared,
including `hill climbing', simulated annealing approaches and
genetic algorithms \citep{Williamson1998}.
These approaches involve the selection and replication of a
discrete number of individuals from a nationally representative list 
such as the SARs. Thus, subsets of individuals are taken from the global
microdataset (geocoded at coarse geographies) and allocated to small areas.
There have been several refinements and applications of the original
ideas suggested by \citet{Williamson1998}, including research reported by
\citet{Voas2000},  \citet{Williamson2002}, and
\citet{Ballas2006}.

\subsection{Deterministic reweighting}
\label{s:det-reweighting}
The methods described in the previous section involve the use of random sampling
procedures or `probabilistic reweighting' \citep{Hermes2012a}. In
contrast, \citet{Ballas2005b} presented an alternative deterministic approach
based on IPF. It is the results of this method, that does not use
random number generators and thus produces the same output with each
run,\footnote{Probabilistic results can also be replicated, by `setting the
seed' of a predefined set of pseudo-random numbers.} that the
integerisation methods presented here take as their starting point.
The underlying theory behind IPF has been described in a number
of papers 
\citep{Deming1940, Mosteller1968, Wong1992}. \citet{Fienberg1970} proves
that IPF converges towards a single solution.

IPF can be used to produce maximum likelihood estimates of spatially
disaggregated conditional probabilities for the individual attributes of interest.
The method is also known as `matrix raking', RAS
or `entropy maximising'
(see \citealp{Johnston1993, Birkin1988, Muller2010, Huang2001a,
Kalantari2008, Jirousek1995}).
The mathematical properties of IPF
have have been described in several papers
(see for instance \citealp{Bishop1975, Fienberg1970, Birkin1988}).
Illustrative examples of the procedure can be found in
\citet{Saito1992}, \citet{Wong1992}
and \citet{Norman1999a}.
\citet{Wong1992} investigated the reliability of IPF and evaluated
the importance of different factors influencing its performance;
\citet{Simpson2005} evaluated methods for improving the performance of
IPF-based microsimulation.
Building on these methods, IPF has been employed by
others to investigate a wide range of phenomena
(e.g.~\citealp{Mitchell2000, Ballas2005c, Williamson2002, Tomintz2008}).

Practical guidance on how to perform IPF for spatial microsimulation
is also available. In an online working paper, \citet{Norman1999a}
provides a user guide for a Microsoft Excel macro that
performs IPF on large datasets. \citet{Simpson2005} provided code snippets of
their procedure in the statistical package SPSS. 
\citet{Ballas2005b} describe the process and how it can be applied to problems
of small area estimation. In addition to these resources, a practical guide to
running IPF in R has been created to accompany this paper.\footnote{This
guide, ``Spatial microsimulation in R: a beginner's guide to
iterative proportional fitting (IPF)'', is available from
\url{http://rpubs.com/RobinLovelace/5089} .}

\subsection{Combinatorial optimisation, IPF and
the need for integerisation}
\label{need}

The aim of IPF, as with all spatial microsimulation methods, is to  match
individual-level data from one source to aggregated data from another.
IPF does this repeatedly, using one constraint variable at a time: each
brings the column and row totals of the simulated dataset closer to
those of the area in question (see \citealp{Ballas2005b} and
Fig.~\ref{fig:IPF-4c} below). 

Unlike combinatorial optimisation algorithms, IPF results in non-integer
weights. As mentioned above, this is problematic for certain applications.
In their overview of methods for spatial microsimulation
\citet{Williamson1998} favoured combinatorial
optimisation approaches, precisely for this reason:
%It should be noted that only integer reweighting schemes are to be considered
``as non-integer weights lead, upon tabulation of results, to fractions of
households or individuals'' (p.\ 791). There are two options
available for dealing with this problem with IPF:
\begin{itemize}
\item Use combinatorial optimisation microsimulation methods instead
\citep{Williamson1998}. However, this can be computationally intensive
\citep{Pritchard2012}.
\item Integerise the weights: Translate the non-integer weights obtained
through IPF into discrete counts of individuals selected from the original
survey dataset \citep{Ballas2005c}.
\end{itemize}
We revisit the second option, which arguably provides the
`best of both worlds': the simplicity and computational speed of deterministic
reweighting and the benefits of using whole cases.
%allows for the
%use of simple IPF models to deterministically provide the best possible fit
%between
% As with any model, spatial microsimulation models are rough approximations
% of an infinitely complex reality and, as with any model, they rely on
% assumptions about how the world works. In this case the founding assumption is
% that the relationships between the constraint variables and the target
%variables
% are the same for individuals in the survey as for individuals in the
% areas under investigation. Using the example of income, its dependence on age
%is
% assumed to remain constant in each area. If young people earn less money in
%the
% survey, an area containing a high proportion of young people will be expected
%to
% have a low income relative to an area containing a low proportion of young
% people, all other factors being equal. The technique makes the assumption
% that relationships between socio-economic variables remain constant over space
% \citep{Cullinan2011}.
% composed of scattered points whose
%characteristics vary are ubiquitous in Geography,
%or `integer reweighting', as it also called \citep{Ballas2005},
%Examples of why integerisation is useful

In summary, IPF is an established method for combining microdata
with spatially aggregated constraints to simulate  target variables whose
characteristics are not recorded at the local level. Intergerisation translates
the real number weights obtained by IPF into samples from the original
microdata, a list of `cloned' individuals for each simulated area.
Integerisation may also be useful conceptually, as it allows
researchers to deal with entire individuals. The next section
reviews existing strategies for integerisation.

\section{Method}
\label{strategies}
Despite the importance of integer weights for dynamic spatial microsimulation,
and the continued use of IPF, there
has been little work directed towards integerisation. It has been noted that ``the
integerization and the selection tasks may introduce a bias in the synthesized
population'' \citep[10]{Muller2010}, yet little work has been done to find out
\emph{how much} error is introduced.

To test each integerisation method, IPF was used to to generate an
array of weights that fit individual-level survey data to
geographically aggregated Census data (see Section \ref{worked-eg}). Five methods for
integerising the results are described, three deterministic and two
probabilistic. These are: `simple rounding', its evolution into the `threshold
approach' and the `counter-weight' method and the probabilistic methods
`proportional probabilities' and finally `truncate, replicate, sample'. TRS
builds on the strengths of the other methods, hence the order in which they are
presented.

The application of these methods
to the same dataset (and their implementation in the same language, R) allows
their respective performance characteristics to be quantified and compared.
Before proceeding to describe the mechanisms by which these integerisation
methods work, it is worth taking a step back, to consider the nature and
meaning of IPF weights.

\subsection{Interpreting IPF weights: replication and probability}
It is important to clarify what we mean by `weights' before proceeding to
implement methods of integerisation: this understanding was central to the
development of the integerisation method presented in this paper.
The weights obtained through IPF are real numbers ranging from 0 to hundreds
(the largest weight in the case study dataset is 311.8). This range
makes integerisation problematic: if the probability of selection is
proportional
to the IPF weights (as is the case with the `proportional probabilities' method),
the majority of resulting selection probabilities can be very low.
This is why the simple rounding method rounds weights up or down to the nearest
integer weight to determine how many times each individual should be
replicated (Ballas et al., 2005a): to ensure replication weights do not differ
greatly from non-integer IPF weights. However, some of the information contained
in the weight is lost during rounding: a weight remainder of 0.501 is treated
the same as 0.999.

This raises the following question: Do the weights refer to the number of times
a particular individual should be replicated, or is it related to the probability
of being selected? The following sections
consider different approaches to addressing this question, and the
integerisation methods that result.

\subsection{Simple rounding}
The simplest approach to integerisation is to convert the non-integer
weights into an integer by rounding. If the decimal remainder to the right of
the decimal is 0.5 or above, the integer is rounded up; if not, the
integer is rounded down.

Rounding alone is inadequate for accurate results, however. As illustrated in
Fig.~\ref{fig:histws} below, the distribution of weights obtained by IPF is likely to
be skewed, and the majority of weights may fall below the critical 0.5 value and
be excluded. As reported by \citet[25]{Ballas2005c}, this results in inaccurate
total populations. To overcome this problem 
\citet{Ballas2005c} developed algorithms to `top up' the simulated
spatial microdata with
representative individuals: the `threshold' and `counter-weight' approaches.

\subsection{The threshold approach}
\citet{Ballas2005c} tackled the need to `top up' the simulated area
populations such that $Pop_{sim} \geq Pop_{cens}$. To do this, an inclusion
threshold ($IT$) is created, set to 1 and then iteratively
reduced (by 0.001 each time), adding extra individuals with
incrementally lower weights.\footnote{A more detailed description of the steps
taken and the R code needed to perform them iteratively can be found in the
Supplementary Information, Section 3.2.}
Below the exit value of $IT$ for each zone, no individuals can be included
(hence the clear cut-off point around 0.4 in Fig.~\ref{fig:threshweights}).
In its original form, based on rounded weights, this approach over-replicates
individuals with high decimal weights.
To overcome this problem, we took the truncated weights as the starting
population, rather than the rounded weights. This modified approach improved the
accuracy of the integer results and is therefore what we refer to when
the `threshold approach' is mentioned
henceforth.\footnote{An
explanation of this improvement can be illustrated by considering an individual
with a weight of 2.99. Under the original threshold approach described by
\citet{Ballas2005c}, this person would be replicated 4 times: three times after
rounding, and then a
fourth time after $IT$ drops below 0.99. With our modified approach they would
be replicated three times: twice after truncation, and again after $IT$ drops
below 0.99. The improvement in accuracy in our tests was substantial, from a TAE
(total absolute error, described below) of 96,670 to 66,762. Because both
methods are equally easy to implement, we henceforth refer only to the superior
version of the threshold integerisation
method.}

The technique successfully tops-up integer populations yet has
a tendency to generate too many individuals for each zone.
This oversampling is due to
duplicate weights --- each unique weight was repeated on
average 3 times in our model --- and the presence of
weights that \emph{are} different, but separated by less than 0.001.
(In our test, the mean number of unique weights falling into
non-empty bins between 0.3 and 0.48 in each area --- the range of values
reached by $IT$ before  $Pop_{sim} \geq Pop_{cens}$ --- is almost two.)

% \begin{enumerate}
%  \item Set the initial value of $IT$ to 1.
% \item If $Pop_{sim} < Pop_{cens}$, run the following loop (if not skip it).
% \item Re-sample or `clone' any individuals whose decimal
% weights\footnote{By `decimal weight', we refer to the
% value of a weight to the right of the decimal point. So, for a weight of 1.8,
% the `decimal weight' is 0.8. Mathematically, the decimal weight (which
% we also refer to as the `weight remainder') can be defined as $w - trunc(w)$
% where the function trunc() removes all information to the right of the decimal.}
% are less than
% $IT$ yet greater than or equal to $IT - x$, where $x$ is a small number to be
% iteratively subtracted from $IT$ (\citet{Ballas2005c} suggest $x = 0.001$; this
% value was also used here).
% \item Recalculate $Pop_{sim}$ with the additional individuals included.
% \item Subtract $x$ from $IT$ to reduce the inclusion threshold for the next
% iteration. If $Pop_{sim}$ is still less than $Pop_{cens}$ return to step 2; if not
% exit.
% \end{enumerate}

% \begin{figure}[h]
%  \centerline{ \includegraphics[width=14 cm]{meths-scat2.pdf}}
%  % meths-scat.pdf: 792x612 pixel, 72dpi, 27.94x21.59 cm, bb=
%  \caption{Comparison of simulated and census populations using rounding and
% threshold techniques for 71 zones (the black line represents x = y).
% }
%  \label{fig:meths-scat}
% \end{figure}

%  The combined impacts
% of these effects are significant (see Section \ref{results}). In practice, an
% `exclusion threshold' could help to overcome this problem,
% although our new strategy overcomes both issues.

\subsection{The counter-weight approach}
An alternative method for topping-up integer results arrived at by simple
rounding was also described by \citet{Ballas2005c}. The approach was labelled
to emphasise its reliance on both counter and a weight variables. Each
individual is first allocated a counter in ascending order of its IPF weight.
The algorithm then tops-up the integer results of simple rounding by iterating
over all individuals in the order of their count. With each iteration the new
integer weight is set as the rounded weight plus the rounded sum of its decimal
weight plus the decimal weight of the next individual, until the desired total
population is reached.\footnote{This process is described in more detail in the
Supplementary Information.} %%% Now do it!

\begin{figure}[t]
 \centerline{ \includegraphics[width=14 cm]{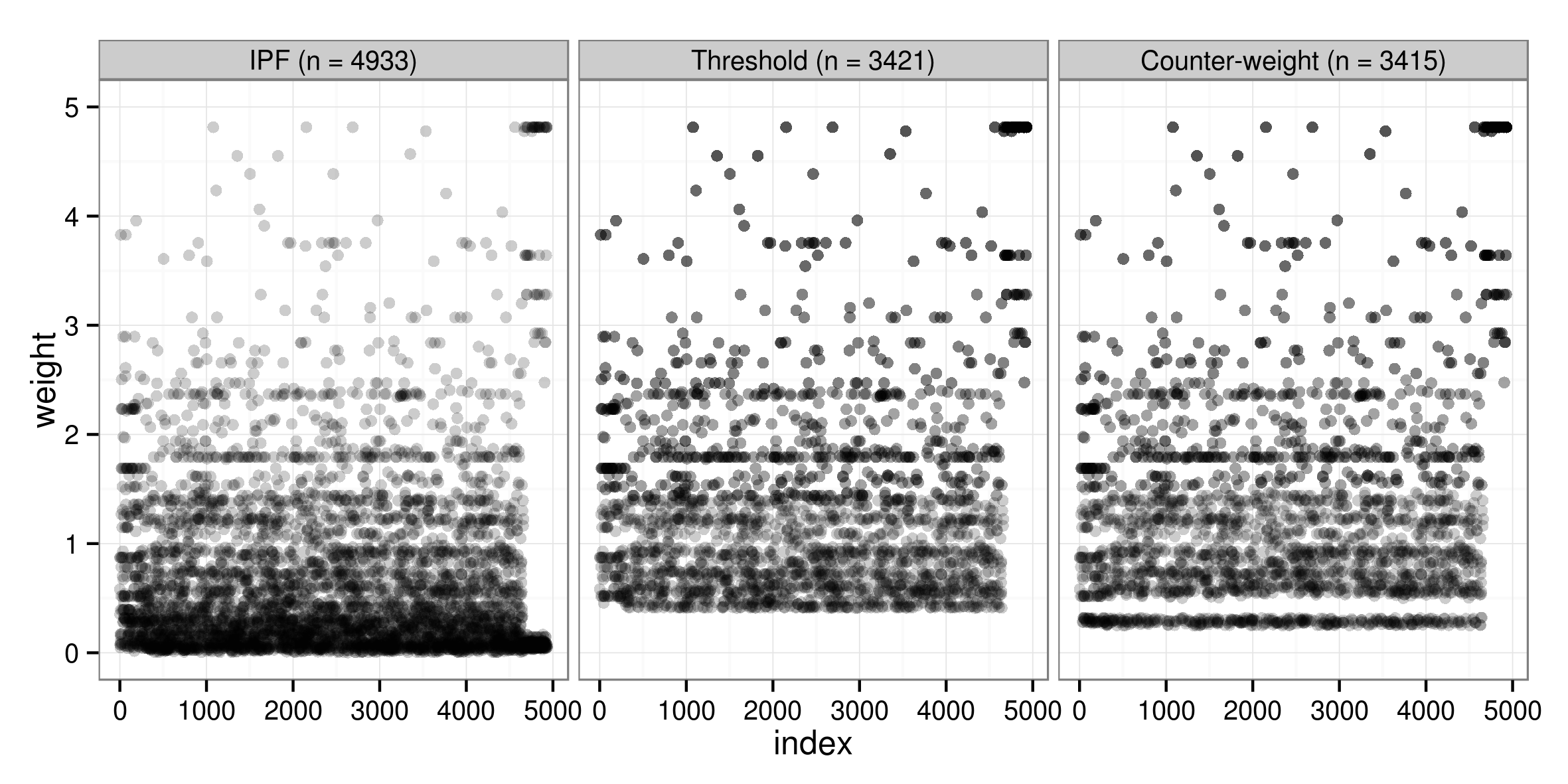}}
 % meths-scat.pdf: 792x612 pixel, 72dpi, 27.94x21.59 cm, bb=
 \caption{Overplotted scatter graph showing the distribution of weights and
replications after IPF in the original survey (left), those selected
by inclusion thresholds for a single area (middle), and those selected
by the counter-weight method (right) for zone 71 in the
example dataset. The lightest points represent individuals who have been
replicated once, the darkest 5 times.}
 \label{fig:threshweights}
\end{figure}

There are two theoretical advantages of this approach: its more accurate
final populations (it does not automatically duplicate individuals with equal
weights as the threshold approach does) and the fact that 
individuals with decimal
weights down to 0.25 may be selected.
This latter advantage is minor, as $IT$ reached below 0.4 in many cases
(Supplementary Information, Fig.~2) --- not far off.
A band of low weights (just above 0.25)
selected by the counter-weight method can be seen in Fig.~\ref{fig:threshweights}.

The total omission of weights below some threshold is problematic for all
deterministic algorithms tested here: they imply that someone with a weight below
this threshold, for example 0.199 in our tests,
has the same sampling probability
as someone with a weight of 0.001: zero! The complete
omission of low weights fails to make use of all the information stored in
IPF weights: in fact, the individual with an
IPF weight of 0.199 is 199 times more representative of the area (in terms of
the constraint variables and the make-up of the survey dataset) than the
individual with an IPF weight of 0.001. Probabilistic approaches to
integerisation ensure that all such differences between decimal weights
are accounted for.

% \begin{figure}[t]
%  \centerline{ \includegraphics[width=14 cm]{Inclusion-hist2.pdf}}
%  % meths-scat.pdf: 792x612 pixel, 72dpi, 27.94x21.59 cm, bb=
%  \caption{Histogram of the inclusion thresholds ($IT$) reached in order
% to `top-up' individuals selected by simple rounding. (n = 71 zones.)}
%  \label{fig:Inclusion-hist}
% \end{figure}

\subsection{The proportional probabilities approach}
This approach to integerisation treats IPF weights as probabilities.
The chance of an individual being selected is proportional to the IPF
weight:
\begin{equation}
 p = \frac{w}{\sum{W}}
\end{equation}
Sampling until $Pop_{sim} = Pop_{cens}$ \emph{with replication} ensures that
individuals with high weights are likely to be repeated several times
whereas individuals with low weights are unlikely to appear.
% Due to the law of rare events, the probability that an individual with
% weight $w$ will appear  $X \in \mathbb{N}$ times in the integerised results follows
% the Poisson distribution:
% \begin{equation}
% p\left( X \right) = \frac{{e^{ - w } w ^X }}{{X!}}
% \end{equation}
The outcome of this strategy is correct from a theoretical perspective,
yet because all weights are treated as
probabilities, there is a non-zero
chance that an individual with a low weight
(e.g.~0.3) is replicated
more times than an individual with a higher weight (e.g.~3.3). (In this
case the probability for any given area is $\sim$ 1\%, regardless of the population size).
Ideally, this should never happen: the individual with weight 0.3 should be replicated
either 0 or 1 times, the probability of the latter being 0.3. The approach
described in the next section addresses these issues.

\subsection{Truncate, replicate, sample}
\label{s:trs}
The problems associated with the aforementioned integerisation strategies
demonstrate the need for an alternative method.
Ideally, the method would build upon the simplicity of the
rounding method, select the correct simulated population size (as attempted by
the threshold approach and achieved by using `proportional probabilities'),
make use of all the information stored in IPF
weights \emph{and} reduce the error introduced by integerisation to a
minimum. The probabilistic approach used in `proportional probabilities'
allows multiple answers to be calculated (by using different `seeds').
This is advantageous for analysis of uncertainty introduced by the
process and allows for the selection of the best fitting result.
Consideration of these design criteria led us to
develop TRS integerisation, which interprets weights as
follows: IPF weights do
not merely represent the probability of a single case being selected. They
also (when above one) contain information about repetition: the two
types of weight are bound up in a single number. An IPF weight of 9, for
example, means that the individual should be replicated 9 times in the
synthetic microdataset. A weight of 0.2, by contrast, means that the
characteristics of this individual should count for only 1/5 of their whole
value in the microsimulated dataset and that, in a representative
sampling strategy, the individual would have a probability of 0.2 of
being selected. Clearly, these are different concepts. As such, the
TRS approach to integerisation isolates the replication and probability 
components of IPF weights at the outset, and then deals with each separately.
Simple rounding, by contrast, interprets IPF weights as inaccurate count data.
The steps followed by the TRS approach are  described in detail below.

\subsubsection{Truncate}
By removing all information to the right of the decimal point, truncation
results in integer values --- integer replication weights that
determine how many times each individual should be `cloned' and placed into the
simulated microdataset. In R, the following command is used: \begin{verbatim}
count <- trunc(w)
\end{verbatim}
% This command is identical to integer division by 1 \begin{verbatim}
% x %/% 1
% \end{verbatim},
where \verb w \ is a matrix of individual weights.  Saving these values (as
\verb count ) will later ensure  that only whole integers are counted. The
decimal remainders (\verb dr ), which vary between 0 and 1, are saved by
subtracting the integer weights from the full weights:\begin{verbatim}
dr <- w - count
\end{verbatim}
This separation of conventional and replication weights provides the
basis for the next stage: replication of the integer weights.

\subsubsection{Replicate}
In spreadsheets, replication refers simply to copying cells of
data and pasting them elsewhere. In spatial microsimulation, the
concept is no different. The number of times a
row of data is replicated depends on the integer weight: an IPF
weight of 0.99, for example, would not be replicated at this stage
because the integer weight (obtained through truncation) is 0.

To reduce the computational requirements of this stage, it is best
to simply replicate the row number (\verb index ) associated with
each individual, rather than replicate the entire row of data. This
is illustrated in the following code example, which appears
within a loop for each area (\verb i ) to be simulated:
\begin{verbatim}
 ints[[i]] <- index[rep(1:nrow(index),count)]
\end{verbatim}

Here, the indices (of weights above 1, \verb index ) are selected
and then repeated. This is done using the function \verb rep() .
The first argument (\verb 1:nrow(index) ) simply defines
the indices to be replicated; the second (\verb count )
refers to the integer weights defined in the previous subsection.
(Note: \verb count \ in this context refers only to the
integer weights above 1 in each area).
 Once the replicated indices have been generated, they
can then be used to look up the relevant characteristics of
the individuals in question.

\begin{figure}[h]
 \centerline{ \includegraphics[width=14 cm]{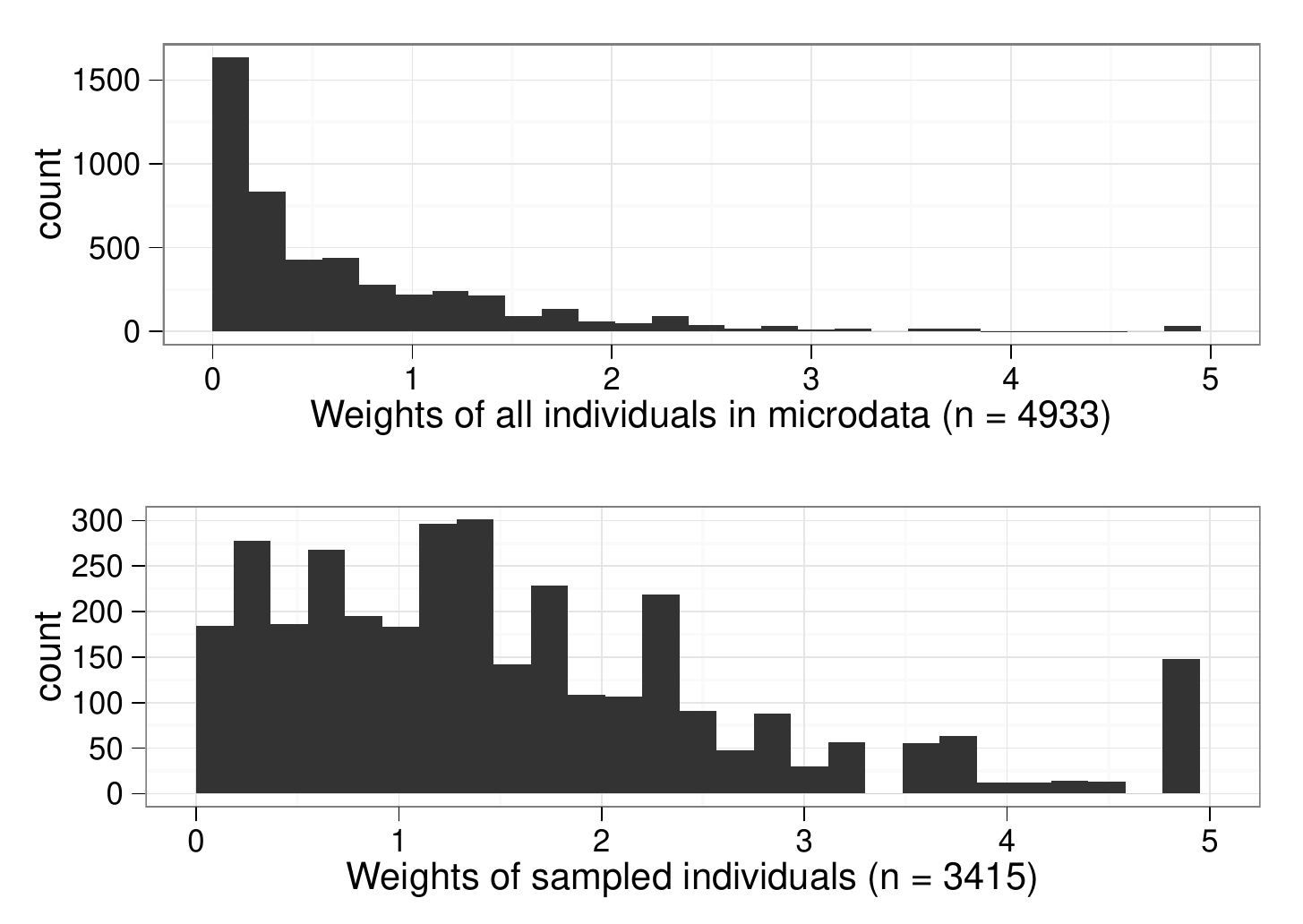}}
 % hist-weights.pdf: 792x612 pixel, 72dpi, 27.94x21.59 cm, bb=0 0 792 612
 \caption{Histograms of original microdata weights (above) and sampled microdata
after TRS integerisation (below) for a single area --- zone 71 in the case study data.}
 \label{fig:histws}
\end{figure}

\subsubsection{Sample}
As with the rounding approach, the truncation and replication stages alone
are unable to produce microsimulated datasets of the correct size. The problem
is exacerbated by the use of truncation instead of rounding: truncation is
guaranteed to produce integer microdataset populations
that are smaller, and in some cases much smaller than the actual (census)
populations. In our case study,
the simulated microdataset populations were around half the
actual size populations defined by the census. This under-selection of
whole cases has the following advantage: when using truncation there
is no chance of over-sampling, avoiding the problem of
simulated populations being slightly too large, as can occur
with the threshold approach.

Given that the replication weights have already been included in steps 1 and 2,
only the decimal weight remainders need to be included. This can be done using
weighted random sampling without replacement. In R, the following function is used:
\begin{verbatim}
  sample(w, size=(pops[i,1] - pops[i,2]), prob= dr[,i])
\end{verbatim}
Here, the argument \verb size \ within the \verb sample \ command is set as the
difference between the known population of each area (\verb pops[i,1] ) and
the size obtained through the replication stage alone (\verb pops[i,2] ). The
probability (\verb prob ) of an individual being sampled is determined by the
decimal remainders. \verb dr \ varies between 0 and 1, as described above.

\begin{figure}[h]
 \centerline{
 \includegraphics[width=12 cm]{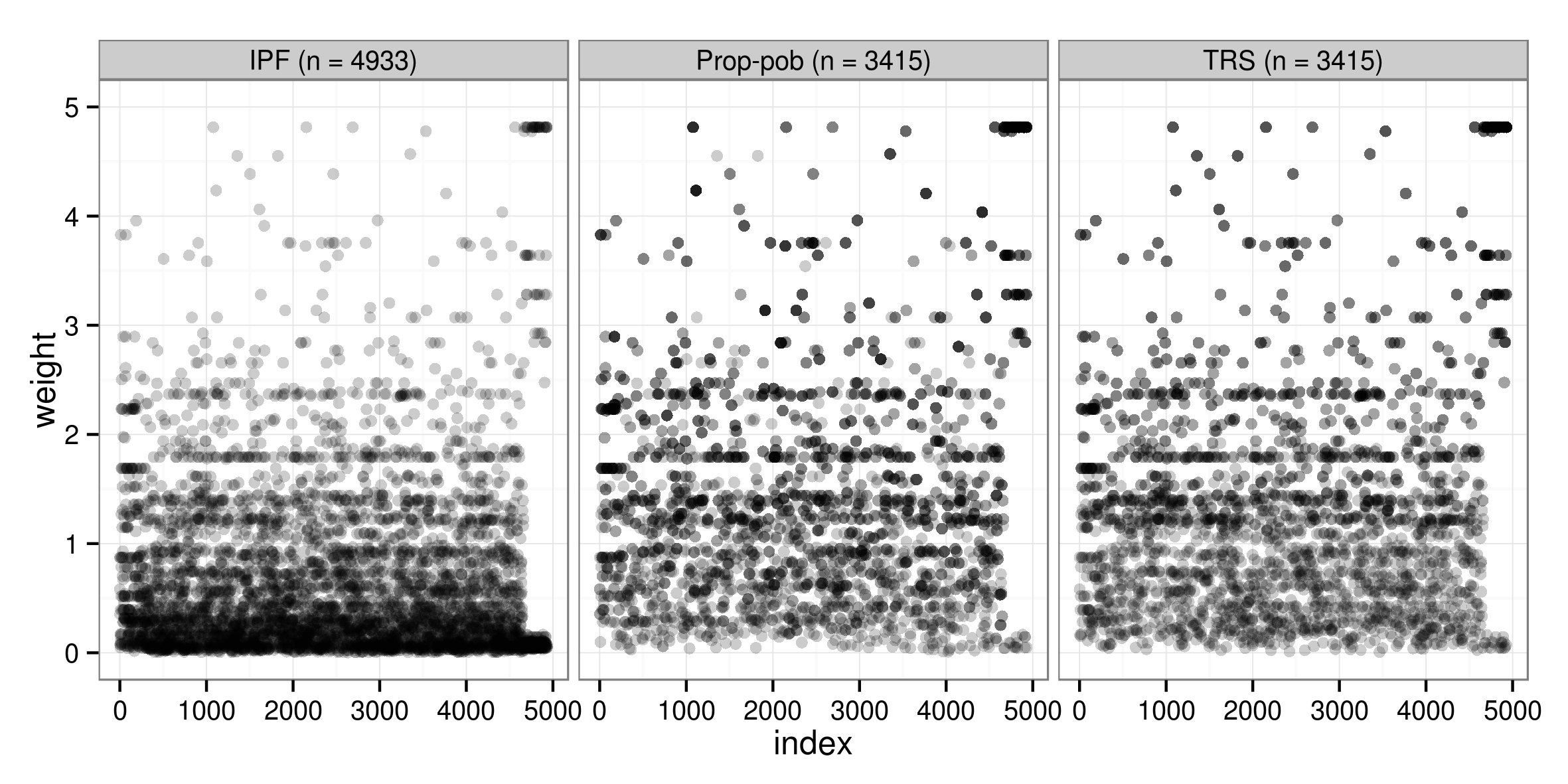}}
 % hist-weights.pdf: 792x612 pixel, 72dpi, 27.94x21.59 cm, bb=0 0 792 612
 \caption{Overplotted scatter graphs of index against weight
 for the original IPF weights (left) and after proportional
 probabilities (middle) and TRS (right) integerisation for zone 71.
Compare with Fig.~\ref{fig:threshweights}.}
 \label{fig:index-weight-TRS}
\end{figure}

The results for one particular area are presented in Fig.~\ref{fig:histws}.
The distribution of selected individuals has shifted to the right, as
the replication stage has replicated individuals as a function of their truncated
weight. Individuals with low weights (below one) still constitute a large
portion of those selected, yet these individuals are replicated fewer times.
After TRS integerisation individuals with high decimal weights are relatively
common. Before integerisation, individuals with IPF weights between 0 and 0.3
dominated. An individual-by-individual visualisation of the Monte Carlo
sampling strategy is provided in Fig.~\ref{fig:index-weight-TRS}. Comparing
this with the same plot for the probabilistic methods
(Fig.~\ref{fig:threshweights}), the most noticeable difference is that the TRS
and proportional probabilities approaches
include individuals with very low weights. Another important difference is average point
density, as illustrated by the transparency of the dots: in
Fig.~\ref{fig:threshweights}, there are shifts near the decimal
weight threshold ($\sim$ 0.4 in this area) on the y-axis.
In Fig.~\ref{fig:index-weight-TRS}, by contrast, the transition is
smoother: average darkness of single dots (the number of replications)
gradually increases from 0 to 5 in both probabilistic methods.

\begin{figure}[h]
 \centerline{
 \includegraphics[width=12 cm]{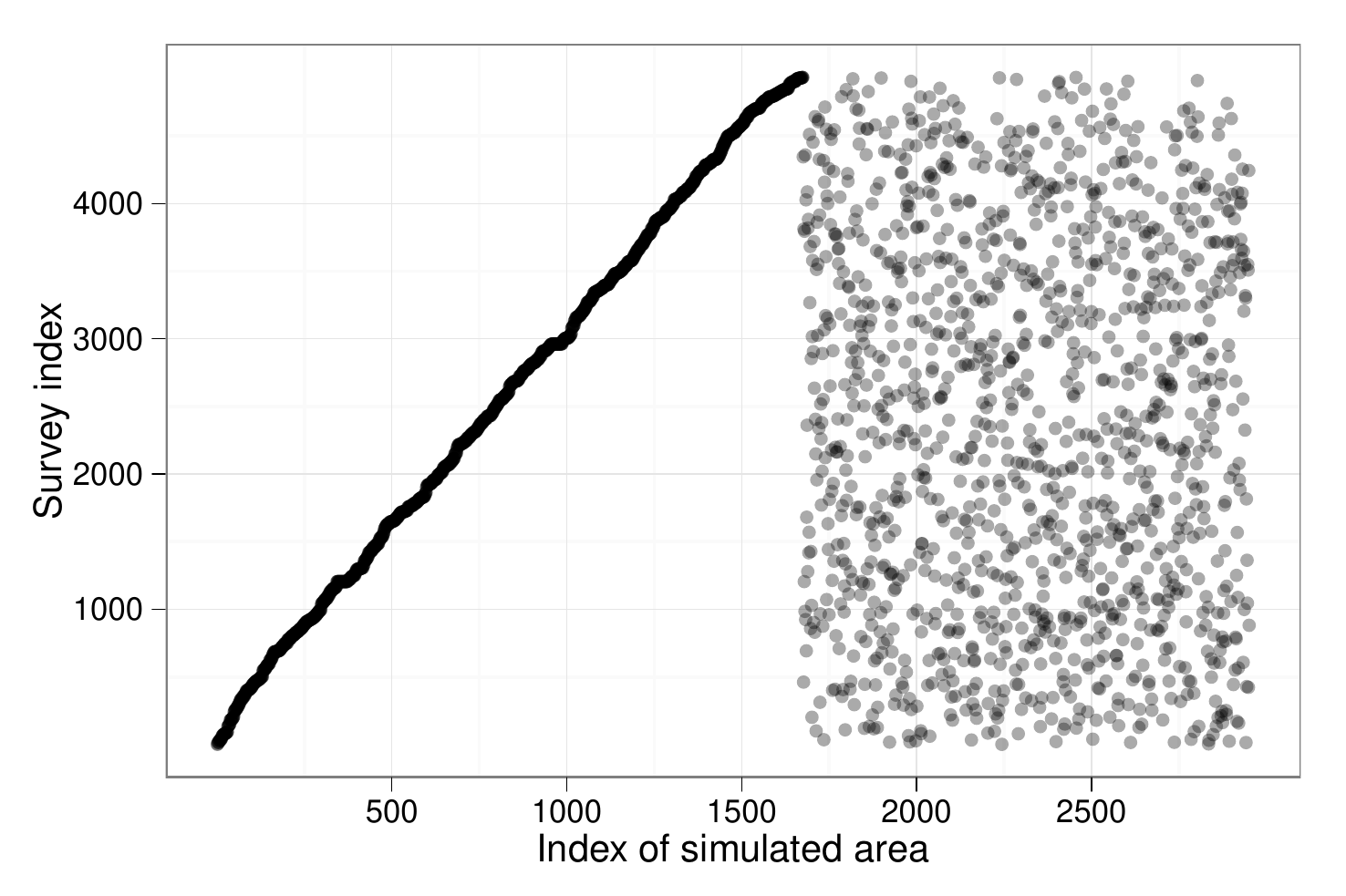}}
 % hist-weights.pdf: 792x612 pixel, 72dpi, 27.94x21.59 cm, bb=0 0 792 612
 \caption{Scatter graph of the index values of individuals in the original
sample and their indices following TRS Integerisation for a single area. }
 \label{fig:index-TRS}
\end{figure}

Fig.~\ref{fig:index-TRS} illustrates the mechanism by which the TRS sampling
strategy works to select individuals. In the first stage (up to x = 1,717,
in this case) there is a linear
relationship between the indices of survey and sampled individuals, as the
model iteratively moves through the individuals, replicating those with
truncated weights greater than 0. This
(deterministic) replication stage selects roughly half of the required population
in our example dataset (this proportion varies from zone to zone).
The next stage is probabilistic sampling
(x = 1,718 onwards in Fig.~\ref{fig:index-TRS}): individuals are selected from
the entire microdataset with selection probabilities equal to weight remainders.

\subsection{The test scenario: input data and IPF}
\label{worked-eg}
% In order to test integerisation techniques, they were tested on a real world
% example. This section serves two functions: to illustrate the utility of
% integerisation techniques when modelling continuous variables using spatial
% microsimulation, and to describe how the integerisation 
% techniques were tested.  

The theory and methods presented above demonstrate how five integerisation
methods work in abstract terms. But to compare them quantitatively a test
scenario is needed. This example consists of a spatial microsimulation model
that uses IPF to model the commuting and socio-demographic characteristics
of economically active individuals in
Sheffield. According to the 2001 Census, Sheffield has a working
population of just over 230,000. The characteristics of these
individuals were simulated by reweighting a synthetic microdataset based on
aggregate constraint variables provided at the medium super output area (MSOA)
level. The synthetic microdataset was created by `scrambling' a subset of the
Understanding Society dataset (USd).\footnote{See
http://www.understandingsociety.org.uk/. To scramble this data, the continuous
variables (see Table \ref{t:data}) had an integer random number (between 10 and
-10) added to them; categorical variables were mixed up, and all other
information was removed.} MSOAs
contain on average
just over 7,000 people each, of whom 44\% are economically active
in the study area; for the less sensitive aggregate constraints, real data were
used. These variables are summarised in Table \ref{t:data}.

\begin{table}[htbp]
\caption{Summary data for the spatial microsimulation model}
\begin{tabular}{lrlll}
\toprule
 & \multicolumn{ 2}{c}{\textbf{Aggregate data}} & \multicolumn{
2}{c}{\textbf{Survey data}} \\
 & \multicolumn{ 2}{c}{71 zones, average pop.: 3077.5} & \multicolumn{
2}{c}{4933 observations} \\ \midrule
Variable & \multicolumn{1}{l}{N. categories} & Most populous  & Mean  &
Most populous \\ \hline
Age / sex  & 12 & Male, 35 to 54 yrs & \multicolumn{1}{r}{40.1} & - \\
Mode  & 11 & Car driver & - & Car driver \\
Distance  & 8 & 2 to 5 km & \multicolumn{1}{r}{11.6} & - \\
NS-SEC  & 9 & Lower managerial & - & Lower managerial \\ \bottomrule
\end{tabular}
\label{t:data}
\end{table}

The data contains both continuous (age, distance) and categorical (mode,
NS-SEC) variables. In practice, all variables are converted into categorical
variables for the purposes of IPF, however. To do this statistical
bins are used.
Table \ref{t:data} illustrates similarities between aggregate
and survey data overall (car drivers being the most popular mode of travel to
work in both categories, for example). Large differences exist between
individual zones and survey data, however: it is the role of iterative
proportional fitting to apply weights to minimize these differences.

\begin{figure}[h]
 \centerline{
 \includegraphics[width=14 cm]{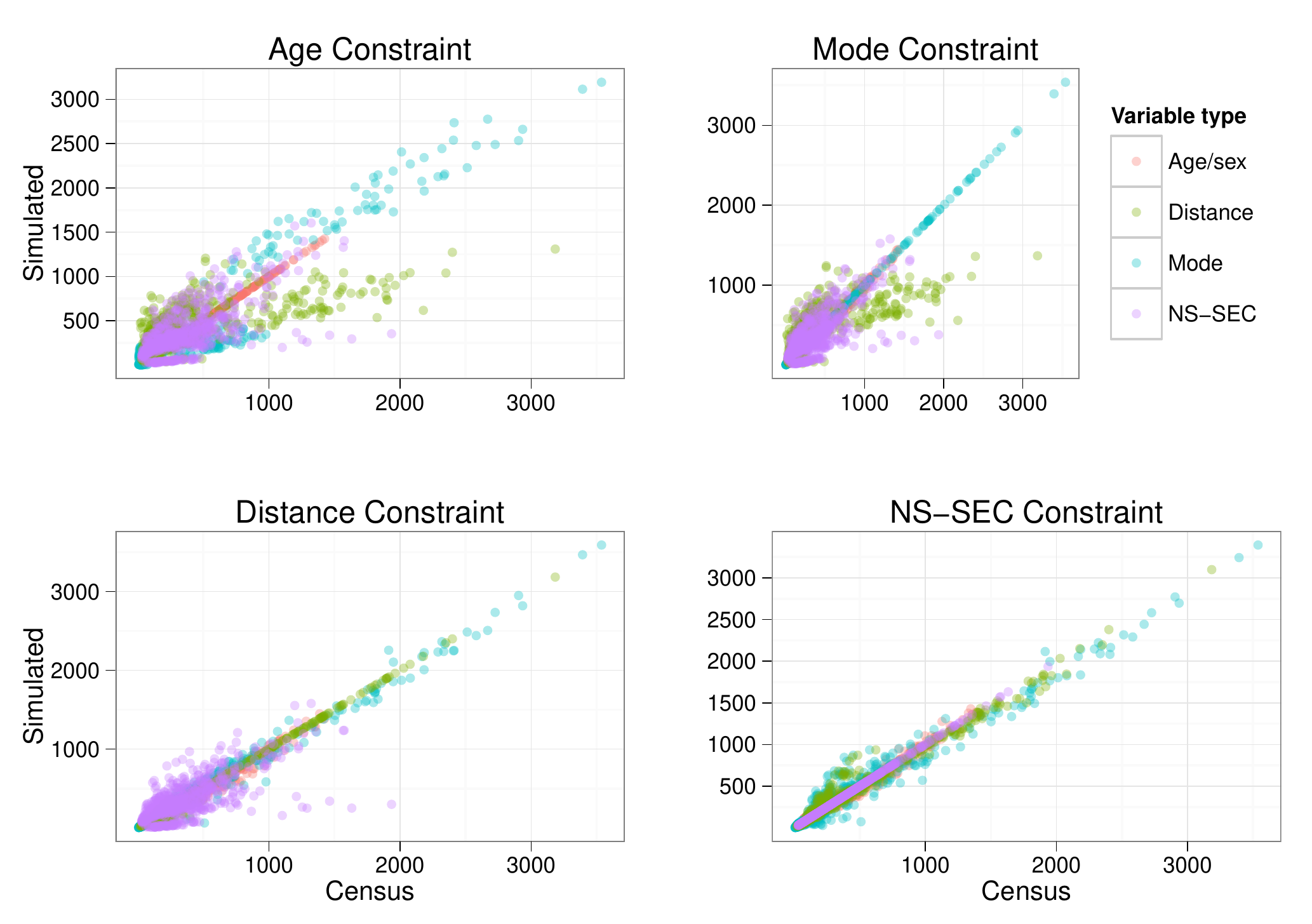}}
 % hist-weights.pdf: 792x612 pixel, 72dpi, 27.94x21.59 cm, bb=0 0 792 612
 \caption{Visualisation of IPF method. The graphs show the iterative improvements
in fit after age, mode, distance and finally NS-SEC constraints were applied
(see Table \ref{t:data}). See footnote 4 for resources on how IPF works.}
 \label{fig:IPF-4c}
\end{figure}

\begin{figure}[h]
 \centerline{
 \includegraphics[width=13 cm]{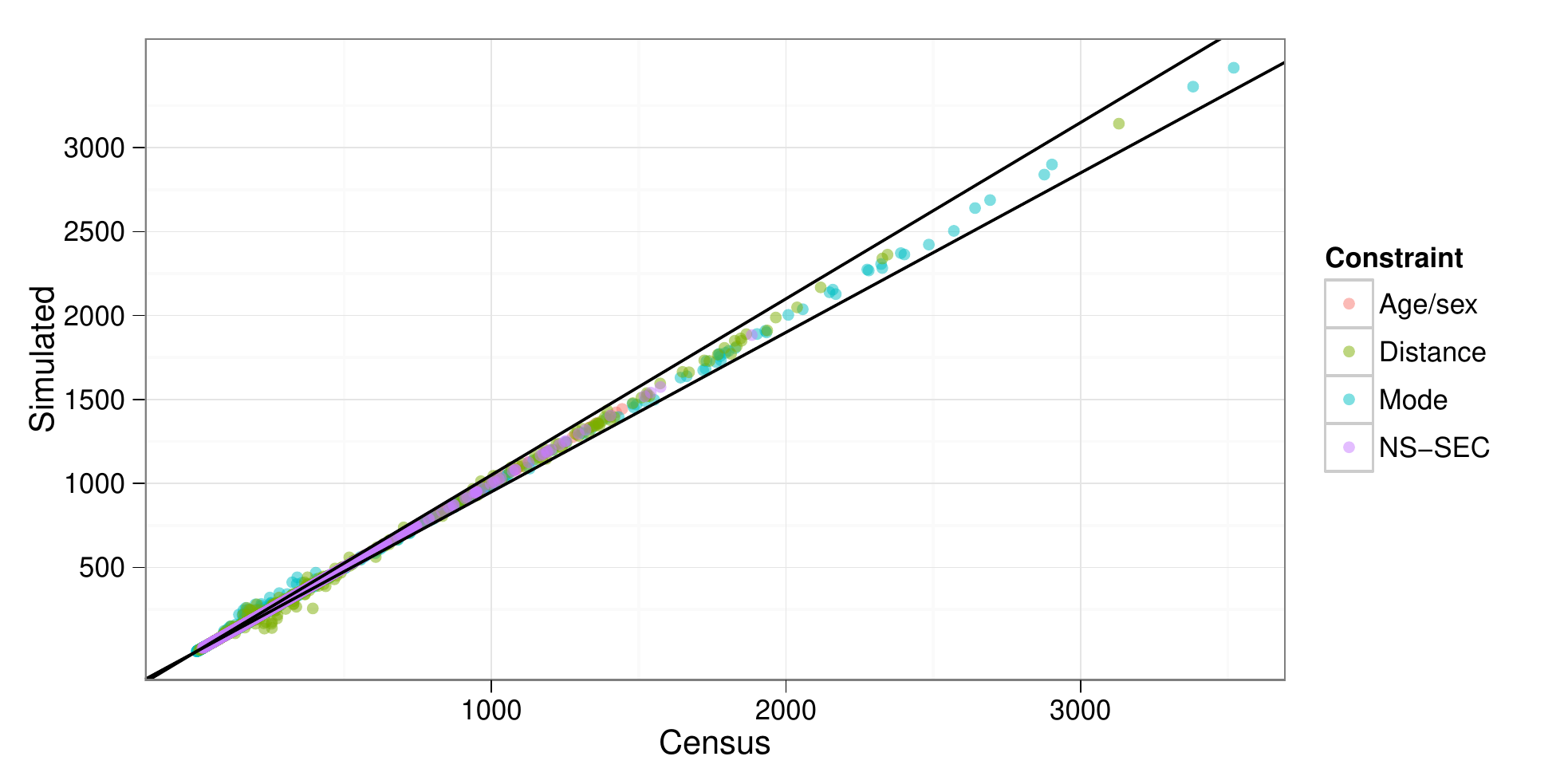}}
 % hist-weights.pdf: 792x612 pixel, 72dpi, 27.94x21.59 cm, bb=0 0 792 612
 \caption{Scatter graph illustrating the fit between census and simulated
aggregates after 20 IPF iterations (compare with Fig.~\ref{fig:IPF-4c}).}
 \label{fig:ipf-scat}
\end{figure}

IPF was used to assign 71 weights to
each of the 4,933 individuals, one weight for each zone. The fit
between census and weighted microdata can be seen
improving after constraining by each of the 40 variables (Fig.~\ref{fig:IPF-4c}).
The process is repeated until an adequate level
of convergence is attained (see Fig.~\ref{fig:ipf-scat}).\footnote{What
constitutes an `adequate' level of fit has not been well defined in the
literature, as mentioned in the next section. In this example, 20
iterations were used.}
The weights were set to an initial value of
one.\footnote{An initial value must be selected for IPF to create new weights
which better match the small area constraints.
It was set to one as this tends to be the average weight value in social surveys
(the mean Understanding Society dataset interview plus proxy individual
cross-sectional weight is 0.986).}
The weights were then iteratively
altered to match the aggregate (MSOA) level statistics, as described in
Section \ref{s:det-reweighting}.

Four constraint variables link the aggregated census data to the survey,
containing a total of 40 categories. To illustrate how IPF works, it is useful
to inspect the fit between simulated and census aggregates before and after
performing IPF for each constraint variable. Fig.~\ref{fig:IPF-4c}
illustrates this process for each constraint. By contrast to existing
approaches to visualising IPF (see \citealp{Ballas2005b}), Fig.
\ref{fig:IPF-4c} plots the results for all variables, one
constraint at a time. This approach can highlight which constraint variables
are particularly problematic. After 20
iterations (Fig.~\ref{fig:ipf-scat}), one can see that distance and mode
constraints are most problematic. This may be because both variables depend
largely on geographical location, so are not captured well by UK-wide
aggregates.

Fig.~\ref{fig:IPF-4c} also illustrates how IPF works: after reweighting for
a particular constraint, the weights are forced to take values such that the
aggregate statistics of the simulated microdataset match perfectly with the
census aggregates, for all variables within the constraint in question.
Aggregate values for the mode variables, for example, fit the census results
perfectly after constraining by mode (top right panel in Fig.
\ref{fig:IPF-4c}). Reweighting by the next constraint disrupts the fit
imposed by the previous constraint --- note the increase scatter of the (blue)
mode variables after weights are constrained by distance (bottom left).

However, the disrupted fit is better than the original. This leads to
a convergence of the weights such that the fit between simulated and known
variables is optimised:
% The fit after each complete iteration can be formally measured in
% absolute and relative terms. The latter case is illustrated in Fig.~\ref{}
Fig.~\ref{fig:IPF-4c} shows that accuracy increases after weights are
constrained by each successive linking variable.
% % %\ref{fig:ints-errors},
% % %which shows how accuracy (measured as the proportion of simulated results
% % %falling beyond 5\% above or below the census value, as illustrated in
% % \ref{fig:4hists}, which shows continual improvement in model fit: the
% % distribution of residuals tend to 0 with each successive iteration.
% % 
% % The results also show, however, that
% % some variables create more error than other --- the results of reweighting by
% % the constraint ``mode'' are always worse than obtained by reweighting by the
% % other variables. This is significant because it means that the final results
% of
% % IPF models may depend more on the order of the constraint variables than on
% the
% % number of iterations. 
% 
% %  \begin{figure}[t]
% %  \centerline{
% %  \includegraphics[width=14 cm]{its-errors.pdf}}
% %  % hist-weights.pdf: 792x612 pixel, 72dpi, 27.94x21.59 cm, bb=0 0 792 612
% %  \caption{Proportion of results that fall beyond 5\% of the census values
% after
% % reweighting by each constraint variable over 10 complete iterations.}
% %  \label{fig:ints-errors}
% % \end{figure}

\section{Results}
\label{results}
This section compares the five previously describe approaches to
integerisation --- rounding, inclusion threshold, counter-weight, proportional
probabilities  and TRS methods. The results are based on the 20$^{th}$
iteration of the IPF model described above. The following metrics of
performance were assessed:
\begin{itemize}
 \item Speed of calculation.
\item Accuracy of results.
\begin{itemize}
 \item Sample size.
%: does the integer microsimulation sample size match the actual population?
\item Total Absolute Error (TAE) of simulated areas.
\item Anomalies (aggregate cell values out by more than 5\%).
\item Correlation between constraint variables in the census and
microsimulated data.
\end{itemize}
\end{itemize}

\begin{figure}[h*]

 \centerline{
 \includegraphics[width=13.5 cm]{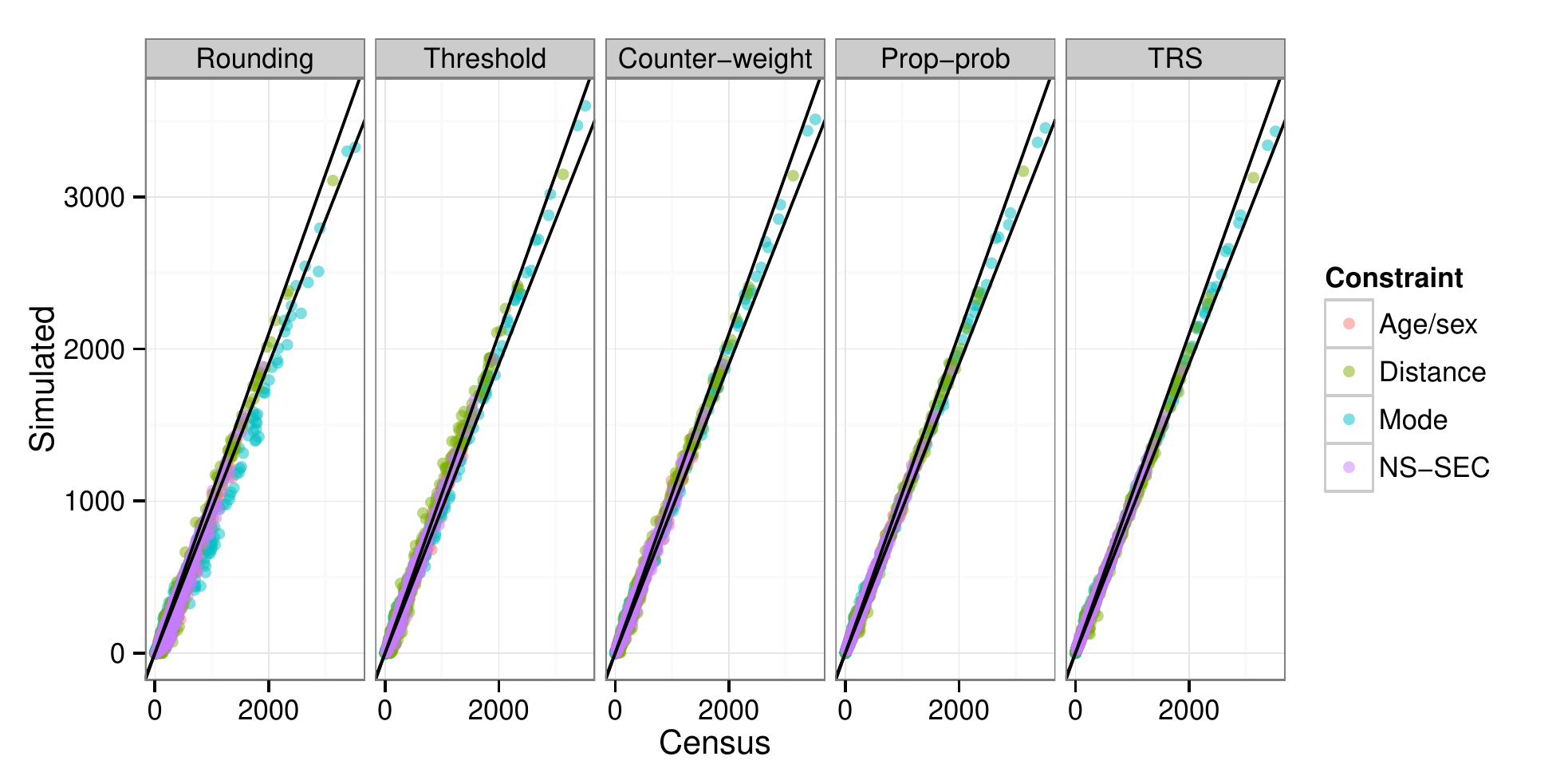}}
 % hist-weights.pdf: 792x612 pixel, 72dpi, 27.94x21.59 cm, bb=0 0 792 612
 \caption{Scatterplots of actual (census) and simulated population totals for
four
integerisation techniques. The black lines represent 5\% error in either
direction. }
 \label{fig:3scat}
\end{figure}

Of these performance indicators accuracy is the most problematic.
Options for measuring goodness-of-fit have proliferated in the last two decades,
yet there is no consensus about which is most appropriate \citep{Voas2001}.
The approach taken here, therefore, is to use a range of measures, the most
important of which are summarised in Table \ref{acc-results} and Fig.
\ref{fig:3scat}.

\subsection{Speed of calculation}
%With modern desktop computers, computational requirements is less frequently 
The time taken for the integerisation of IPF weights was measured on an Intel
Core i5 660 (3.33 GHz) machine with 4 Gb of RAM running Linux 3.0.
The simple rounding method of integerisation was unsurprisingly the fastest, at
4 seconds. % Add c-w time
In second and third place respectively were the proportional probabilities
and TRS approaches, which took a a couple of seconds longer for a single
integerisation run for all areas.
Slowest were the inclusion threshold and counter-weight techniques, which took
three times longer than simple rounding. To ensure representative results for
the probabilistic approaches, both were run 20 times and the result with the
best fit was selected. These imputation loops took just under a minute.

The computational intensity of integerisation may
be problematic when processing weights for very large
datasets, or using older computers. However, the results must be placed in
the context of the computational requirements of the IPF process itself. For the
example described in Section \ref{worked-eg}, IPF took approximately
30 seconds per iteration and 5 minutes for the full 20 iterations.

\subsection{Accuracy}

In order to compare the fit between simulated microdata and the zonally
aggregated linking variables that constrain them, the former must first be
aggregated by zone. This aggregation stage allows the fit between linking
variables to be compared directly (see Fig.~\ref{fig:3scat}). More
formally, this aggregation allows goodness of fit to be calculated using a range
of metrics \citep{Williamson1998}. We compared the accuracy of integerisation
techniques using 5 metrics:
\begin{itemize}
\item Pearson's product-moment correlation coefficient ($r$).
\item Total and standardised absolute error (TAE and SAE),
\item Proportion of simulated values falling beyond 5\% of the actual values,
% This is the standard measure
% of correlation and requires no further explanation here \citep{Rodgers1988}.
\item The proportion of Z-scores significant at the 5\% level.
\item Size of the sampled populations, 
\end{itemize}

The simplest way to evaluate the fit between simulated and census results
was to use Pearson's $r$, an established measure of association \citep{Rodgers1988}.
The $r$ values for all constraints were 0.9911, 0.9960, 0.9978,
0.9989 and 0.9992 for rounding, threshold, counter-weight, proportional
probabilities and TRS
methods respectively. IPF alone had an $r$ value of 0.9996. These correlations
establish an order of fit that can be compared to other metrics.

TAE and SAE are crude yet effective measures of overall
model fit \citep{Voas2001}. TAE has the additional advantage of
being easily understood:
\begin{equation}
 TAE = \sum\limits_{ij}|U_{ij} - T_{ij}|
\end{equation}
where U and T are the observed and simulated values for each linking variable
($j$) and each area ($i$).
% Change 34 (symbol: % Change): converted TEA to TAE
SAE is the TAE divided by the total
population of the study area. TAE is sensitive to the number of people
within the model, while SAE is not. The latter is seen by \citet{Voas2001} as
``marginally preferable'' to the former: it allows cross-comparisons between
models of different total populations \citep{Kongmuang2006-thesis}.

\begin{table}[]
\caption{Accuracy results for integerisation techniques.*}
\small{
\begin{center}
\begin{tabular}{llrrrr}
\toprule
Method & Variables & \multicolumn{1}{l}{TAE} & \multicolumn{1}{l}{SAE (\%)} &
\multicolumn{1}{l}{E $>$ 5\% (\%)} & \multicolumn{1}{l}{$Z{m}^{2}$ (\%)} \\
\midrule
IPF & Age/sex & 9 & 0.0 & 0.0 & 0.0 \\
 & Distance & 4874 & 2.3 & 13.7 & 4.9 \\
 & Mode & 4201 & 2.0 & 6.4 & 4.2 \\ 
 & NS-SEC & 0 & 0.0 & 0.0 & 0.0 \\ 
\textbf{} & \textbf{All} & \textbf{9084} & \textbf{3.1} & \textbf{4.5} & \textbf{2.1} \\ 
\midrule
Round- & Age/sex & 26812 & 12.5 & 81.5 & 39.8 \\ 
 ing& Distance & 31981 & 14.9 & 80.1 & 65.1 \\ 
 & Mode & 30558 & 14.2 & 81.4 & 48.9 \\ 
 & NS-SEC & 27493 & 12.8 & 76.5 & 57.1 \\ 
 & \textbf{All} & \textbf{116844} & \textbf{13.6} & \textbf{80.1} & \textbf{51.3} \\ 
\midrule
Thresh- & Age/sex &  11076 & 5.1 & 49.2 & 8.1 \\ 
 old& Distance & 27146 & 12.6 & 82.4 & 57.7 \\
 & Mode & 14770 & 6.9 & 68.6 & 33.9 \\
 & NS-SEC & 13770 & 6.4 & 55.2 & 24.1 \\
 & \textbf{All} &  \textbf{66762} & \textbf{7.8} & \textbf{62.5} & \textbf{28.7}
\\ 
\midrule
Counter- & Age/sex & 10242 & 4.8 & 47.7 & 6.6 \\
weight& Distance &  17103 & 8.0 & 70.2 & 39.3 \\ 
 & Mode &  10072 & 4.7 & 60.4 & 21.6 \\
 & NS-SEC & 11798 & 5.5 & 49.6 & 17.1 \\
 & \textbf{All} & \textbf{49215} & \textbf{5.7} & \textbf{56.1} & \textbf{19.6}
\\
\midrule

Propor- & Age/sex & 9112 & 4.2 & 48.0 & 3.1 \\ 
tional & Distance & 8740 & 4.1 & 47.4 & 10.4 \\ 
proba- & Mode & 8664 & 4.0 & 60.8 & 9.0 \\ 
bilities & NS-SEC & 7778 & 3.6 & 37.6 & 3.3 \\ 
 & \textbf{All} & \textbf{34294} & \textbf{4.0} & \textbf{49.0} & \textbf{6.2} \\
\midrule
TRS &Age/sex & 5424 & 2.5 & 27.9 & 0.4 \\ 
 & Distance & 10167 & 4.7 & 48.8 & 16.4 \\ 
 & Mode & 7584 & 3.5 & 56.1 & 6.7 \\ 
 & NS-SEC & 5687 & 2.6 & 24.9 & 1.1 \\ 
 & \textbf{Total} & \textbf{28862} & \textbf{3.4} & \textbf{39.2} & \textbf{5.5} \\ 
\bottomrule
\end{tabular}
\end{center}
}
\label{acc-results}
\begin{tablenotes}
      \footnotesize
      \item * The probabilistic
results represent the best fit (in terms of TAE) of 20 integerisation runs
with the pseudo-random number seed set to 1000 for replicability --- see
Supplementary Information.
    \end{tablenotes}
\end{table}

The proportion of values which fall beyond 5\% of the actual values is a simple
metric of the quality of the fit. It implies that getting a perfect fit is not
the aim, and penalises fits that have a large number of outliers. The precise
definition of 'outlier' is somewhat arbitrary (one could just as well use 1\%).

The final metric presented in Table \ref{acc-results}
is based on the Z-statistic, a standardised measure of
deviance from expected values, calculated for each cell of data. We use $Zm$, a
modified version of the Z-statistic which is a robust measure of fit for each
cell value \citet{Williamson1998}. The measure of
fit is appropriate here as it takes into account absolute, rather than just
relative, differences between simulated and observed cell count:
\begin{equation}
 Zm_{ij} = (r_{ij} - p_{ij}) \Bigg/ \left(\frac{p_{ij}(1 -
p_{ij}))}{\sum\limits_{ij}U_{ij}}\right)^{1/2}
\end{equation}

where

\begin{center}
\begin{math}
  p_{ij} = \frac{U_{ij}}{\sum\limits_{ij}U_{ij}} \qquad and \qquad r_{ij} =
\frac{T_{ij}}{\sum\limits_{ij}U_{ij}}
\end{math}
\end{center}

To use the modified Z-statistic as a measure of overall model fit, one simply
sums the squares of $zm$ to calculate $Z{m}^{2}$. This measure can handle
observed cell counts below 5, which chi-squared tests cannot \citep{Voas2001}.

The results presented in Table \ref{acc-results} confirm that \emph{all}
integerisation methods introduce
some error. It is reassuring that the comparative accuracy is the same across
all metrics. Total absolute error (TAE), the simplest goodness-of-fit
metric, indicates that discrepancies between simulated and census data increase
by a factor of 3.2 after TRS integerisation, compared with raw
(fractional) IPF weights.\footnote{In the case of a sufficiently diverse input
survey dataset, IPF would be able to find the perfect solution: TAE would be 0
and the ratio of error would not be
applicable.}
Still, this is a
major improvement on the simple rounding, threshold and counter-weight
approaches to integerisation presented by \citet{Ballas2005c}: these increased
TAE by a factor of 13, 7 and 5 respectively.
The improvement in fit relative to the proportional probabilities method
is more modest. The proportional probabilities method increased TAE by a factor
of 3.8, 23\% more absolute error than TRS.

The differences between the simulated and actual populations ($Pop_{sim} -
Pop_{cens}$) were also calculated for
each area. The resulting differences are summarised in Table 5, which
illustrates that the counter-weight and two probabilistic methods resulted
in the correct population totals for every area. Simple rounding and threshold
integerisation methods greatly underestimate and slightly overestimate the
actual populations, respectively.

\begin{table}[h*]
\begin{center}
\caption{Differences between census and simulated populations.}
\vspace{0.25 cm}
\begin{tabular}{lrrr}
\toprule
Metric & \multicolumn{1}{l}{Rounding} & \multicolumn{1}{l}{Threshold} &
\multicolumn{1}{l}{Others (CW, PP, TRS)} \\ \midrule
Mean & -372 & 8 & 0  \\
Standard deviation & 88 & 11 & 0 \\
Max & -133 & 54 & 0 \\
Min & -536 & 0 & 0 \\
Oversample (\%) & -13 & 0.3 & 0 \\
\bottomrule
\end{tabular}
\end{center}
\label{t:pops}
\end{table}

\section{Discussion and conclusions}
\label{discuss}
The results show that TRS integerisation outperforms the other methods of
integerisation tested in this paper.
At the aggregate level, accuracy
improves in the following order: simple rounding,
inclusion threshold, counter-weight, proportional probabilities and, most
accurately, TRS. This order of preference remains unchanged, regardless of which
(from a selection of 5) measure of goodness-of-fit is used. These results
concur with a finding derived from theory --- that ``deterministic rounding of
the counts is not a satisfactory integerization''
\citep[p.~689]{Pritchard2012}.
Proportional probability and TRS methods clearly provide more accurate
alternatives.

An additional advantage of the probabilistic TRS and proportional probability
methods is that correct population sizes are guaranteed.\footnote{Although
the counter-weight method produced the correct population sizes in our tests, it
cannot be guaranteed to do so in all cases, because of its reliance on simple
rounding: if more weights are rounded up than down, the population will be too
high. However, it can be expected to yield the correct population in cases
where the populations of the areas under investigation are substantially
larger than the number of individuals in the survey dataset.}
In terms of speed of calculation, TRS also performs well. TRS takes marginally
more time than simple rounding and proportional probability methods,
but is three times quicker than the threshold and counter-weight
approaches. In practice, it seems that integerisation processing time is
small relative to running IPF over several iterations. Another
major benefit of these non-deterministic methods is that probability
distributions of results can be generated, if the algorithms are run multiple
times using unrelated pseudo-random numbers. Probabilistic methods could
therefore enable the uncertainty introduced through integerisation to be
investigated quantitatively
\citep{Beckman1996, Rubin1987} and subsequently illustrated using error bars.

Overall the results indicate that TRS is superior to the
deterministic methods on many levels and introduces less error than the
proportional probabilities approach.
We cannot claim that TRS is `the best' integerisation strategy available though:
there may be other solutions to the problem and different sets of test weights
may generate different
results.\footnote{Despite these caveats, the order of accuracy
identified in this paper is expected to hold in most cases.
Supplementary Information (Section 4.4), shows the same order of
accuracy (except the threshold method and counter-weight
methods, which swap places) resulting from the integerisation of a
different weight matrix.
}
The issue will still present a
challenge for future researchers considering the use of IPF to generate sample
populations composed of whole individuals: whether to use deterministic or
probabilistic methods is still an open question (some may favour
deterministic methods that avoid psuedo-random numbers, to ensure
reproducibility regardless of the software used), and the question of whether
combinatorial optimisation algorithms perform better has not been addressed.

Our results provide insight into the advantages and disadvantages of
five integerisation methods and guidance to researchers wishing to
use IPF to generate integer weights: use
TRS unless determinism is needed or until superior alternatives (e.g.~real small
area microdata) become available. Based on the code and example datasets
provided in the Supplementary Information, we encourage others to use, build-on
and improve TRS integerisation.

A broader issue raised by the this research, that requires further
investigation before answers emerge, is `how do the integerised results of IPF
compare with combinatorial optimisation approaches to spatial microsimulation?'
Studies have compared non-integer results of IPF with
alternative approaches \citep{Smith2009, Ryan2009, Rahman2010, harland2012}.
However, these have so far failed to compare like with like: the integer results
of combinatorial approaches are more useful (applicable to more types of
analysis) than the non-integer results of IPF. TRS thus offers a
way of `levelling the playing field' whilst minimising the error introduced to
the results of deterministic re-weighting through integerisation.

In conclusion, the integerisation methods presented in this paper make 
integer results accessible to those with a working knowledge of IPF. TRS
outperforms previously published methods of integerisation. As such, the
technique offers an attractive alternative to combinatorial
optimisation approaches for applications that
require whole individuals to be simulated based on aggregate data.
\section{Acknowledgements}
Thanks to: Mark Green, Luke Temple, David Anderson and Krystyna Koziol for proof reading and suggestions; to
Eveline van Leeuwen for testing the methods on real data and improving the code; and
to the anonymous reviewers for constructive comments.
This research was funded by the Engineering and Physical Sciences Research council (EPSRC) the via the E-Futures
Doctoral Training Centre.

\bibliographystyle{model2-names}
\bibliography{library.bib}

\end{document}

% \title{Supplementary information: a user manual for the integerisation of IPF weights using R}
\title{Supplementary information: a user manual for the integerisation of IPF
weights using R} \author{Robin Lovelace\\ Geography Department
\\University of Sheffield,\\ Sheffield,\\ United Kingdom,\\ S10 2TN\\ }

\date{\today}

\maketitle

This worked example demonstrates how the methods described in the paper
```Truncate, replicate, sample': a method for creating integer weights for
spatial microsimulation'' (Lovelace and Ballas, 2013) were
conducted in R, a free and open source object-orientated statistical programming
language. An introduction to performing iterative proportional fitting (IPF)
in R is provided in a separate document.\footnote{This document is
titled ``Spatial microsimulation in R: a beginner's guide to
iterative proportional fitting (IPF)'', is available from
\url{http://rpubs.com/RobinLovelace/5089} .
A larger project, aimed at optimising R code for IPF applications
can be found at \url{https://github.com/Robinlovelace/IPF-performance-testing} .}
This worked example focuses on
methods for converting the results into integer weights,
with reference to the code that
accompanies this guide.  The main aims are to:
\begin{enumerate}
 \item Introduce R as a user friendly and flexible tool to perform spatial
microsimulation and analyse the results;
\item Demonstrate the replicability of the results described in the paper;
\item  Encourage unrestricted access to code within the microsimulation
community. It is
hoped that this will:
\begin{itemize}
 \item Enhance transparency, repeatability and knowledge transfer within the
field;
\item Allow others to use, test
and further develop existing work, rather than starting from nothing each time,
and;
\item Allow other researchers to critically assess the four integerisation
methods presented in the paper --- named simple round, the threshold approach,
proportional probabilities and truncate, replicate, sample (TRS) ---
so they can be improved.
\end{itemize}
\end{enumerate}
This worked example can be used in different ways, depending on one's aims.
The first section shows how the necessary files can be downloaded and loaded
into R. Section 2 (Running the Spatial Microsimulation Model) is
linked to aim 1, and shows how the microsimulation model, which
forms the foundation of the analysis presented in the paper,
 works.  Section 3 (Integerisation in R) is linked to aim 2,
and demonstrates how to run each type of integerisation, and display some of the
results. Finally, Section 4 (Adapting the Model),
illustrates how the code constituting the spatial microsimulation model
and integerisation techniques can be adapted
to different constraint variables for areas with different population sizes.
Aim 3 can be met throughout: we encourage other researchers to experiment
with and re-use our code, citing this work where appropriate.
To do this, the first stage is to download the dataset and load it into R.

\section{Downloading and loading the files into R}
\label{downl}
Throughout this example, we assume that the data has been
\htmladdnormallink{downloaded}{https://dl.dropbox.com/u/15008199/ints-public.zip}
and extracted to a folder titled `ints-public' onto your desktop and that
\htmladdnormallink{R}{http://www.r-project.org/} is installed on your computer.
To download R, visit the project's homepage and follow the instructions.
For new R users, it is recommended that a introductory text is acquired and
referred to throughout. A range of excellent introductory guides are available
online at http://cran.r-project.org/other-docs.html.

To access the files, unzip the file titled `ints-public.zip', which is available
online in the supplementary
data.\footnote{From here: https://dl.dropbox.com/u/15008199/ints-public.zip}
A list of the folders and files contained
within the folder `ints-public' is provided in \cref{eg-files}.

\begin{table}
\footnotesize{
\caption{Files and folders contained in the worked example folder}
\begin{tabular}{p{3cm}p{3cm}p{6cm}}
\toprule
\textbf{Folder} & \textbf{File} & \textbf{Description} \\
\midrule
\multirow{6}{3cm}{etsim: Spatial microsimulation model based on IPF} & Four
*.csv files, \hspace{2cm} e.g. age-sex.csv & Constraint variables at MSOA level. Based on
scramble census data. \\ \cmidrule{ 2- 3}
 & etsim.R & The spatial microsimulation model resulting
in non-integer weights \\ \cmidrule{ 2- 3}
 & cons.R & R script to read constraint
variables \\ \cmidrule{ 2- 3}
 & plotsim.R & R script to plot the model output \\ \cmidrule{ 2- 3}
 & USd.cat.r & Script to re-aggregate the results \\ \cmidrule{ 2- 3}
 & Usd.RData & Scrambled survey data based on the
Understanding Society dataset \\ \cmidrule{ 2- 3}
its: a subfolder  & etsim1.R etc. & Additional IPF iterations \\ \midrule
\multirow{8}{3cm}{R: folder containing the R code to integerise the weights
generated through IPF} & int-meth1-round.R & Simple rounding method
\\ \cmidrule{ 2- 3}
 & int-meth2-thresh.R & Deterministic threshold method \\ \cmidrule{ 2- 3}
 & int-meth3-cw.R & Counter-weight method \\ \cmidrule{ 2- 3}
 & int-meth4-pp.R & Proportional probability method \\ \cmidrule{ 2- 3}
 & int-meth4-pp-many-runs.R & Many runs of PP method \\ \cmidrule{ 2- 3}
 & int-meth5-TRS.R & Truncate, replicate sample method \\ \cmidrule{ 2- 3}
 & int-meth5-TRS-many-runs.R & Many runs of TRS \\ \cmidrule{ 2- 3}
 & Analysis.R & Analysis of the results of integerisation\\ \cmidrule{ 2- 3}
& outputs.R & Code to generate some results comparing
the 3 integerisation methods \\ \cmidrule{ 2- 3}
 & Plotting-ints.R & Plotting commands \\ \cmidrule{ 2- 3}
 & Iteration-20.RData & Results of the 20th
iteration of IPF \\ \midrule
OA-eg  & see section 4 & Adapted model (for alternative inputs) \\
\bottomrule
\end{tabular}
\label{eg-files}
}
\end{table}

To use the data, the first stage is to set the working directory. Find
out the current working directory using the command \verb getwd() \ from
the R command line.
Correctly setting the working directory will allow quick access the files of the
microsimulation model and a logical place to save the results. The command
\verb list.files() \ is used to check the contents of the working
directory from within R. Assuming the
folder `ints-public' has been extracted to the desktop in a Windows 7 computer
with the user-name `username', type the following into the R command line
interface and press enter to set the working directory (change `username' to
your personal user name or
retype the path completely if the folder was extracted elsewhere):
\begin{verbatim}
setwd("C:/Users/username/Desktop/ints-public/etsim")
\end{verbatim}
To run the model, one can simply type the following (warning: this may take
several minutes, so entering the code block-by-block is recommended):
\begin{verbatim}
source("etsim.R") 
\end{verbatim}

If the aim is to understand
how the method works, we recommend opening the script files using a
text editor and sending the commands to R block by block. This can be done by
copying and pasting blocks of code into the R command line.
Alternatively a graphical user interface such as Rstudio can be used.
In both cases, running the code contained in \verb etsim.R \ should take around
one minute on modern computers, depending on the CPU.  This will result in a
number of objects being loaded onto your R session's workspace. These objects
are listed by the command
\begin{verbatim}
ls()
\end{verbatim}
and can be referred to by name. The constraint variables,
for example, can be summarised using the command:
\begin{verbatim}
 summary(all.msim)
\end{verbatim}
R objects can also be loaded directly, having been saved from previous sessions.
The command:
\begin{verbatim}
 load("iteration-20.RData")
\end{verbatim}
for example, when run in the working directory `R', will load the results of
the IPF model
results after 20 iterations. This may be useful for users who want to move
straight to
integerisation, without running the IPF model first. Referring to file-names in
R can be made easier using the auto-complete capabilities of some R editors.
Rstudio, for example, allows auto-completion of file-names and R objects
(see \htmladdnormallink{RStudio's}{http://rstudio.org/} website for more
information).

\section{Running the spatial microsimulation model}
\label{msim}
The code for running the spatial microsimulation model is
contained within `etsim.R' the folder entitled `etsim'
--- see Table \ref{eg-files}.
As with all R script files, the contents of this
file can be viewed using any
text editor. With an R console running, R's reaction to each chunk of code can
be seen by copy and pasting the script code line-by-line. This should give some
indication of how the model works,
and which parts take most time to process.
Note that R accepts input from external files. Within `etsim.R',
this technique was used to reduce the number of lines of code and
make the model modular.
The constraint variables, for example, are read-in using the following command,
that is contained within the main etsim.R script file:
\begin{verbatim}
  source(file="cons.R")
\end{verbatim}
As before, this simply sends the commands contained within the file to R,
line by line, but without displaying the results until the script has finished.

It is good practice to provide comments within the code,
so that others can see what is going on.
In R, this is done using the hash symbol (\verb # ). Anything
following the hash is ignored by R, until a new line is formed.

% To run the microsimulation model, simply send the contents of `etsim.R'
% into R. Make sure that the working directory has been set correctly,
% so R can access the files it needs. It is possible to send the entire
% contents of the file to R at this stage. However, this may take a few
% minutes to process, as the IPF process over 10 iterations.
% To run the model for just one iteration, do not add the command
% \verb source(file="etsim2.R") . This external file re-runs the
% IPF code again, creating the model's second iteration.

Once the first iteration of the entire model has run, you can check to
see if the model has worked by analysing the objects that R has created.
The raw weights are saved as `weights0', `weights1', etc. The number of each
set of weights corresponds to the constraint which was applied. All of
`weights0' are set to 1 in the first iteration
(the initial condition). The object `weights5' represents the cumulative weight so far,
after the weights have been constrained by all 4 constraint variables.

\begin{figure}[h]
 \centerline{ \includegraphics[width=9 cm]{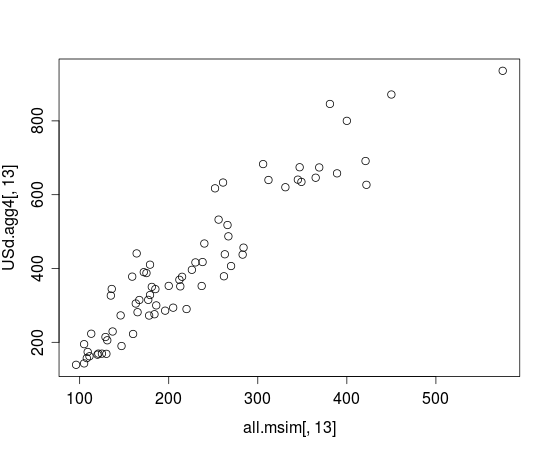}}
 % hist-weights.pdf: 792x612 pixel, 72dpi, 27.94x21.59 cm, bb=0 0 792 612
 \caption{Diagnostic scatter plot to check if the model has worked.}
 \label{mfh}
\end{figure}

The simulated zonal aggregates are stored in objects labeled `USd.agg'
(this stands for `Understanding Society dataset, aggregated'),
from the original value (the summary results of the survey data)
to `USd.agg4' (after fitting for the fourth constraint).
A good first indication of whether the model has
worked is to compare `USd.agg4' with `all.msim' (the latter being
the census aggregates). This can be done by using the command:
\begin{verbatim}
 head(USd.agg4)
\end{verbatim}
and running the same command for `all.msim'. The command
\verb head() \ simply displays the first 5 rows or elements of an R object, to
get a feel for what it looks like. (The meaning of any command can be prefixing
the command name by ``?''. In this case \verb ?head() \ would be used.) To make
the comparison more
interesting, one can plot the results. Try the following:
\begin{verbatim}
plot(all.msim[,13],USd.agg4[,13])
\end{verbatim}
The `\verb [, ' part of this command means ``all rows within''; the
`\verb 13] ' part means ``in column 13''. In this model, column 13
is the variable ``mainly works from home'' (\verb "mfh" ). This
can be established using \verb names(all.msim) , to identify the
variables contained within the dataframe.
If the plot looks the same as as that illustrated below (Fig~.\ref{mfh}),
the model has worked.

\section{Integerisation in R}
\label{ints}
The code used for integerisation of the results of IPF reweighting are kept in
a separate folder, entitled `R' (see Table \ref{eg-files}). As before, navigate
to this folder using a modified version of the following command, this time
navigating to the folder `R':
\begin{verbatim}
setwd("C:/Users/username/Desktop/ints-public/R")
\end{verbatim}

As always, it is worth opening the script files in a text editor or within a
dedicated R development environment such as RStudio. This will
allow the commands to be seen in context and experimented with.

\subsection{Simple rounding}

The following section describes the code contained within the file
`int-meth1-round.R'. Open the file and send its content to
R line by line. The aim of this script is to round the IPF weights (calculated
in the previous section) up or down and then select
individuals accordingly. Once the IPF weights have been loaded using the
\verb load() \ command, the new weights are created using the following
command:\footnote{Here `i20.w5' refers to the weights that emerge after
the 4$^th$ constraint of the 20$^th$ iteration.
Any weights can be used. For example `i1.w5', if loaded into the R
workspace, represents the weights after a single iteration of IPF.}
\begin{verbatim}
intp <- round(i20.w5)
\end{verbatim}
In the following line of code, the decimal remainders are saved by subtracting
the rounded weights from the original weights. Note that each new set of data is
given a name, ready to be referred back to later:
\begin{verbatim}
deci <- i20.w5 - intp # Decimal weights
\end{verbatim}

Before running a loop to select individuals based on their rounded weights, we
created a number of R objects to be used during integerisation.
Of note, the object \verb pops \ is a dataframe for saving data about the
population of each zone that are calculated while the loop is in operation. It
has the same number of rows as there are areas in the constraint table
(\verb 1:nrow(all.msim) ). The columns are bound together using \verb cbind() .
The contents of this object are updated with each iteration, allowing the
results for different methods to be compared directly.

In order to perform calculations on one zone at a time, a loop is used:
\begin{verbatim}
for (i in 1:nrow(all.msim)){ ... }
\end{verbatim}
The commands contained within the curly brackets are performed many times, once
for each area. The index lists the
row name of all individuals within the area \verb i \ who have a rounded
weight above 0 --- \verb which(intp[,i]>0) . The corresponding weight is referred
to by
\verb intp[which(intp[,i]>0),i] . The final list of individuals is saved by
replicating the integer weights the same number of times as the integer value of
the rounded weights:
\begin{verbatim}
ints[[i]] <- index[rep(1:nrow(index),index[,2])]
\end{verbatim}
The replication command \verb rep()  is used in this instance to replicate
the individuals who are `cloned' more than once. Note the use of double square
brackets. This is used to refer to objects (dataframes in this case) that are
part of a
list. Because the matrix of rounded IPF
weights (`intp')
has indexes that correspond to the original survey data, we can extract their
characteristics by simply referring to the previously defined index:
\begin{verbatim}
intall[[i]] <- USd[ints[[i]],]
\end{verbatim}
Finally, the results are aggregated by converting the raw data into the
categories of the constraint variables ---
using \verb source("area.cat.R") --- and then
summing columns to provide zone-wide counts for each category:
\begin{verbatim}
 intagg[i,]   <- colSums(area.cat)
\end{verbatim}
This same procedure is followed for each of the remaining 2 integerisation
methods. The defining features of each are outlined below.

\subsection{The inclusion threshold approach}
The starting point of this method is an incomplete simulated population
of integer results (the length of which is defined as $Pop_{sim}$).
The 5 steps of the threshold approach are as follows:
\begin{enumerate}
 \item Set the initial value of the inclusion threshold $IT$ to 1.
\item If the simulated population is too small ($Pop_{sim} < Pop_{cens}$),
run the following loop (if not skip it).
\item Re-sample or `clone' any individuals whose decimal
weights\footnote{By `decimal weight', we refer to the
value of a weight to the right of the decimal point. So, for a weight of 1.8,
the `decimal weight' is 0.8. Mathematically, the decimal weight (which
we also refer to as the `weight remainder') can be defined as $w - trunc(w)$
where the function trunc() removes all information to the right of the decimal.}
are less than
$IT$ yet greater than or equal to $IT - x$, where $x$ is a small number to be
iteratively subtracted from $IT$ (Ballas et al.~ (2005) --- in SimBritain: a
spatial microsimulation approach to population dynamics --- suggest $x = 0.001$; this
value was also used here).
\item Recalculate $Pop_{sim}$ with the additional individuals included.
\item Subtract $x$ from $IT$ to reduce the inclusion threshold for the next
iteration. If $Pop_{sim}$ is still less than $Pop_{cens}$ return to step 2; if not
exit.
\end{enumerate}

The script file `int-meth2-thresh.R' replicates these steps
in two main loops, each
iterating over the areas whose populations are being simulated. The first
is identical to that of the simple rounding approach, (except in
this case the IPF weights are truncated, not
rounded)\footnote{This
differs from the original implementation in the original
SimBritain paper by Ballas et al. (2005): they used rounded weights
as the starting point. However, after trying both methods, we
found that beginning with truncated integer results leads to far less
error introduced during integerisation. This is because topping up
after simple rounding would lead an individual with a weight of 2.99 to be
replicated 4 times: three times during rounding and once more as the
inclusion threshold dips below
0.99.}
and saves the resulting
microdata as a list of vectors, each containing row names of
individuals from the Understanding Society dataset (see \verb ints[[i]] \ for
area \verb i ).

The second loop adds additional individuals to those contained in \verb ints \
for each area, by gradually reducing the inclusion threshold.
This is done in a third loop which is nested within the second.
Note that the value of the threshold (\verb wv , for weighting variable
--- equivalent to $IT$, as described above)
is set to 1 outside this third loop. This is done so that the threshold
is reduced from one iteration to the next within the loop:
\begin{verbatim}
 wv <- wv - 0.001
\end{verbatim}
Note also that this third loop is initiated by the command
\verb while() , instead of the command \verb for() \ used for
the previous loops. This is because the number of iterations
performed by the first two loops
is fixed (to the number of areas), while the number of
iterations in this one is determined by the threshold at which
the sample population is greater than or to equal the census
population:
\begin{verbatim}
while (length(ints[[i]]) < pops[i,1]){
   \end{verbatim}
Within this sub-loop additional individuals are added whose
 weights are between \verb wv \ and \verb wv $- 0.001$:
\begin{verbatim}
ints[[i]] <- c(ints[[i]], which(dr[,i] < wv & dr[,i] >= wv - 0.001))
\end{verbatim}
Here, the command \verb c() \ appends the additional individuals
to those already saved. After the while loop exits, the population
and aggregate data for each area are saved, as with the simple rounding
method.

\begin{figure}
 \includegraphics[width = 14 cm]{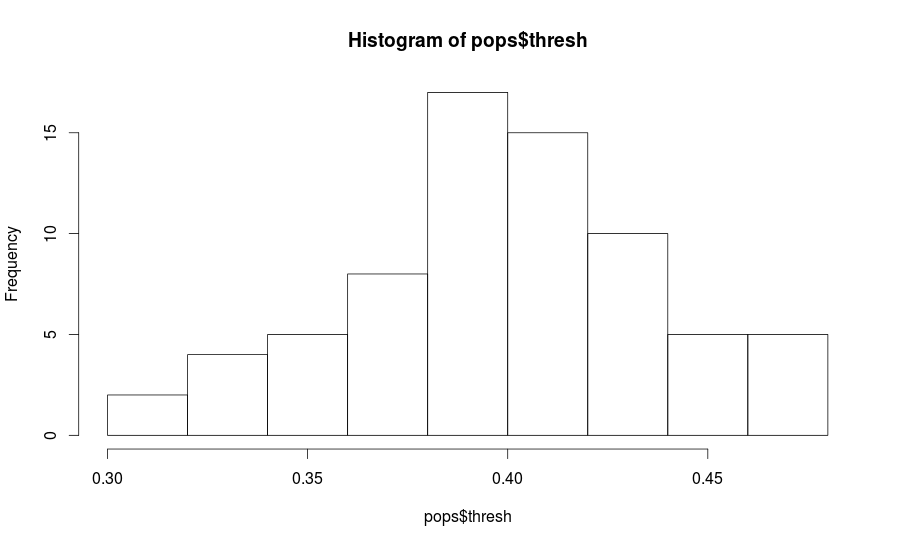}
\caption{Histogram of the lowest value reached by $IT$ (or `wv' in the code)
for all areas during the threshold approach.
}
\label{hist-thresh}
\end{figure}

To analyse the threshold reached for each area, this information is saved as
for each area within the main loop:
\begin{verbatim}
pops$thresh[i] <- wv
\end{verbatim}

This information can be subsequently analysed, e.g.~to investigate the distribution
of thresholds reached (Fig.~\ref{hist-thresh} --- This plot was produced by the following
command:
\verb hist(pops$thresh) ). A similar process is used to save
information about the
exit point of the counter-weight algorithm.

\subsection{The counter-weight method}
The counter-weight method is similar to the threshold algorithm:
it begins with a crude integerisation strategy (in this case
simple rounding, not truncation as described above --- this starting
point was found to lead to more accurate results), and then tops up
each area with additional individuals that depend on their decimal
weights.

The process can be summarised in the following 4 steps:
\begin{itemize}
 \item Sort the IPF weights in ascending order:
      \begin{verbatim}sweights <- sort(i20.w5[,j], index.return = T)$x
      \end{verbatim}
      and save their order for future reference:
      \begin{verbatim}ord <- rank(i20.w5[,j], ties.method="first")
       
      \end{verbatim}
 \item If the total population is too small, top up the results for
 each individual by the rounded sum of their decimal weight plus
 the decimal weight of the next individual in the sorted vector of
 weights:
 \begin{verbatim}
  if(sum(iweights) < round(sum(sweights))){
      iweights[i] <- iweights[i] + round(dweights[i] + dweights[i+1])
      e[j] <- i
    }
 \end{verbatim}

 \item Update the integer weight vector for each area,
 including topped-up individuals, and re-order:
 \begin{verbatim}
    intp[,j] <- iweights # but the order is wrong
  intp[,j] <- intp[ord,j]
 \end{verbatim}
 \item Convert these weight vectors into a list of individuals
 with replicated weights leading to replicated (cloned individuals):
 \begin{verbatim}
  for(i in 1:j){
  index <- cbind((which(intp[,i]>0)) # generates index
                 ,intp[which(intp[,i]>0),i]) # integers)
  ints[[i]] <- index[rep(1:nrow(index),index[,2])] #clone
  pops$pcounter[i] <-  length(ints[[i]]) # save integer dataindividuals
  intall[[i]] <- USd[ints[[i]],] # Pulls all other data from index
  source("area.cat.R")
  intagg[i,]   <- colSums(area.cat)
}
 \end{verbatim}
Finally, the aggregate results for this integerisation method are
saved as with previous methods, in this case as intagg.cw:
\begin{verbatim}
 intagg.cw <- intagg
\end{verbatim}

\end{itemize}

\subsection{Proportional probability method}
The script file `int-meth3-pp.R' also contains two loops.
The first simply creates proportional weights for each
individual-zone combination using the following command:
\verb i20.w5[,i] ~\verb / ~\verb sum(i20.w5[,i]) .
This is the code equivalent of the following equation:
\begin{equation}
 p = \frac{w}{\sum{W}}
\end{equation}
The result (saved as \verb prop.weights[,i] ) is
used in the second loop as the selection probability
for each individual.

The second loop contains three main parts.
First, individuals are randomly selected from the USd
dataset, with probability set as follows:
\begin{verbatim}
 prob=prop.weights[,i] 
\end{verbatim}
(Note that here we are sample \emph{with} replacement ---
\verb replace=T ). Second, the population of the integerised
sample is saved. Third, as with all integerisation methods,
the command \verb source("area.cat.R") ~ is run to extract the
additional information about individual from the Understanding Society
dataset, based solely on their index. The results are saved as
\verb intagg.prop . The next stage is to run the TRS integerisation
method.

\subsection{TRS integerisation in R}
The final method is contained in the script file `Int-meth4-TRS.R'.
It involves weight truncation, replication of integerised weights, and
sampling based on the decimal remainders. Of these steps, sampling is the only
one which requires detailed attention here: the others have already been
described. Suffice to say that integer weights are generated by the command
\verb x%/%1 , which is synonymous with the command \verb trunc(x) . Note that
the command \verb round() \ was used for integerisation in the
simple rounding and threshold integerisation methods.

The population following truncation is guaranteed to be less than the census
population as no rounding up occurs.
This differs from the simple rounding and threshold approaches, and ensures that
there will always be a difference between census and simulated results.
The challenge is to fill the difference:
\begin{verbatim}
popstrs[i,1] - popstrs[i,2]
\end{verbatim}
where \verb popstrs[,1] \ is the census population and \verb popstrs[,2] \ is
the simulated population based on truncated weights.
The command:
\begin{verbatim}
 sample()
\end{verbatim}
allows an exact number of rows to be selected to make up the difference.
The first argument of the command is the vector from which the sample is taken.
The second is the sample size. For our purposes, the vector is the row names
of all individuals from the survey. This vector is referred to by the command
\verb which(i20.w5[,i]>-1) , which means ``all individuals with weights greater
than $-1$, for area \verb i '', i.e.~all individuals. The size is the difference
between census and simulated population sizes for the area in question
(as defined above).

So far so good, but the sample strategy is simple random, meaning
that probabilities will be equally assigned to all rows, unless stated.
This is where the decimal weights --- the `conventional weight' components
of the IPF weights --- come into play. Conventional weights can be used to
determine the probability of an individual being selected.

The final argument used, therefore, is the probability of selection
(\verb prob=... ).
The decimal weights are calculated in-situ by subtracting the integer weights
from the actual weights:
\begin{verbatim}
prob = i20.w5[,i]-i20.w5[,i] %/% 1))\end{verbatim}
As with the previous methods, the loop finishes by extracting the full
survey data from the survey dataset, and saving the aggregate level results:
\begin{verbatim}
 intagg.trs[i,]   <- colSums(area.cat)
\end{verbatim}

After the script files associated with all four integerisation methods have been
run, the aggregate results are saved in R objects entitled \verb intagg.round ,
\verb intagg.thresh \ and \verb intagg.trs . These results form the basis of
the integerisation method performance comparison presented in the paper, and
can be replicated using the file `Analysis.R' (Table \ref{eg-files}).

\section{Adapting the model}
So far the model has been used on a single case study. For the techniques
showcased here to be truly useful, they must be be applicable to a wide range
of situations. This section therefore illustrates how to adapt the model to
simulate the individuals living in Output Areas (which contain around 300 people
or $\sim$100 employed people,
20 times smaller than the Medium Super Output Areas used up until now), using different
constraints and a different (smaller) survey dataset from which individuals are
to be extracted.

\subsection{Setting-up the constraint variables}
In order to show the model's flexibility, 3 new constraint variables were used:
\begin{itemize}
 \item Hours worked per week
\item Marital status
\item Housing tenure of home
\end{itemize}
These variables are available in both aggregate form for small areas,
and from the Understanding Society dataset. The aggregate data
can be downloaded by UK academics from the Casweb census data
portal. The raw data (named `hrs\_{}worked.csv' `marital\_{}status.csv' and `tenancy.csv')
is read into R and cleaned by the commands contained in the script file
`cons.R' within the folder `OA-eg'. The comments in this script file should
explain most of the commands, which read the .csv files and remove superfluous
variables. In one case (tenancy) the variables are also manipulated such that
the category `other' is the sum of three other variables:

\begin{verbatim}
 ten$other <- ten$other + ten$council + ten$assoc
\end{verbatim}

The reason for modifying the data in this way is so that the constraint
data match the individual-level survey data. Also, the USd is a huge dataset
(50994 rows by 1322 columns, contained in a 90 Mb file). Dropping unneeded
information makes the data more manageable.

The script to load and subset the USd data is contained in the file
`load.R' (also in the folder `OA-eg'). For confidentiality reasons
the original data is not provided; the steps taken to process the USd
dataset into a form ready for spatial microsimulation should be
applicable to any survey dataset (the R package `foreign' may be
used to load unusual data types as an R object).
The steps taken here should be fairly
self-explanatory, based on the names of the commands and the comments.
Although the script has been set-up to process the USd survey, in
anticipation of running IPF constrained by the three
constraint variables mentioned above, it would be possible to modify
`load.R' to accept different input survey datasets and subset the data
for different constraints.

The data is also simplified to match available constraint categories in `load.R'.
To provide one example, the USd variable for married status --- `pmarstat' ---
contains 14 categories, many of which can be merged. To
ensure the categories of the survey data matched the census constraints
(5 marriage status categories), the following command was used:
\begin{verbatim}
 levels(Und.sub$mas <- s[sample(nrow(s), size=500),]rstat) <- c(
	      rep("other",5), "single", "married", "single",
            "separated", "divorced", "widowed", rep("other",3))
\end{verbatim}

After running both `load.R' and `cons.R' we are left with four R objects in the
workspace:\footnote{Due to data confidentiality, the full USd dataset
cannot be provided. However, the data that results from
`load.R' has been saved as `oa-data.RData' in the example folder.}
`s', the survey micro-level dataset and `hrs', `mar' and `ten' ---
the three constraint variables.

\subsection{Modifying the spatial microsimulation model}
The script that runs the spatial microsimulation model in the
previous example is called `etsim.R'. In order for it to use new constraint
variables it must be modified. These modifications (which maintain the
original structure and semantics of the original script) can be seen by
comparing `etsim.R' contained within the `OA-eg' folder against
the file of the same name contained within the folder `etsim'.
The following points summarise the changes made:
\begin{itemize}
 \item Add or remove constraints and loading functions depending on the input
data. In this case, for example, the input survey dataframe `s' is too large
relative to the average size of the zones under investigation
(nrow(s) = 1678, more than 10 times greater the average size
of individuals in Output areas ---  $\sim$ 100). Therefore a simple
random sample is taken to reduce the number of rows to 500:
\begin{verbatim}
 s <- s[sample(nrow(s), size=500),]
\end{verbatim}
\item Alter the file `USd.cat.r' so to convert the survey dataframe `s'
into a wide data frame whose dimensions match `all.msim'.
This involves converting categorical variables into binary (1 or 0)
using the subsets. Females who work more than 48 hours per week,
for example, are allocated the value of 1 in the appropriate column
using the following command:
\begin{verbatim}
 s.cat[which(s$jbhrs >= 49 & s$sex=="female"),12] <- 1
\end{verbatim}
\item Names of the R objects referred to are changed to reflect the
new input data. The object `USd', for example, is renamed as `s'.
\end{itemize}

\subsection{Integerisation of the new results}
The integerisation scripts must also be modified slightly
to accept the new input data.
Therefore the files `int-meth1-round.R' to `int-meth4-TRS.R'
described in Table \ref{eg-files} have been
altered. The changes we need to account
for include the new name of the weights (i1.w4
instead of i20.w5 in this case --- only iteration of the
new model has been run for brevity) and, again, the renaming of
the survey to `s' from `USd' in the original files.
It is recommended that differences in the R scripts for integerisation between
files in the folder `R' and those (with the same file names)
in the folder `OA-eg' are identified to understand how the methods can
be generalised to accept any weighted input data.

\subsection{Results}
To confirm that the TRS method advocated in the paper is also the
most accurate when it is used on different input data,
a basic analysis script has been
compiled (`basic-analysis.R' with the folder `OA-eg').
These commands calculate the correlation
between the simulated and census data at the aggregate data
and illustrate the results.
The results demonstrate that the TRS method is also more accurate than
the others for these new constraints, as expected.
The level of correlation rises
(from 0.948 through 0.976, 0.975, and 0.981 to 0.987) for the threshold,
rounding, counter-weight, proportional probabilities and TRS methods
respectively. Note, the order of accuracy is the same as the same
as presented in paper which this Supplementary Information accompanies,
except for the counter-weight method performs worse than the
inclusion threshold approach with the new input datasets.

\begin{figure}
 \includegraphics[width = 14 cm]{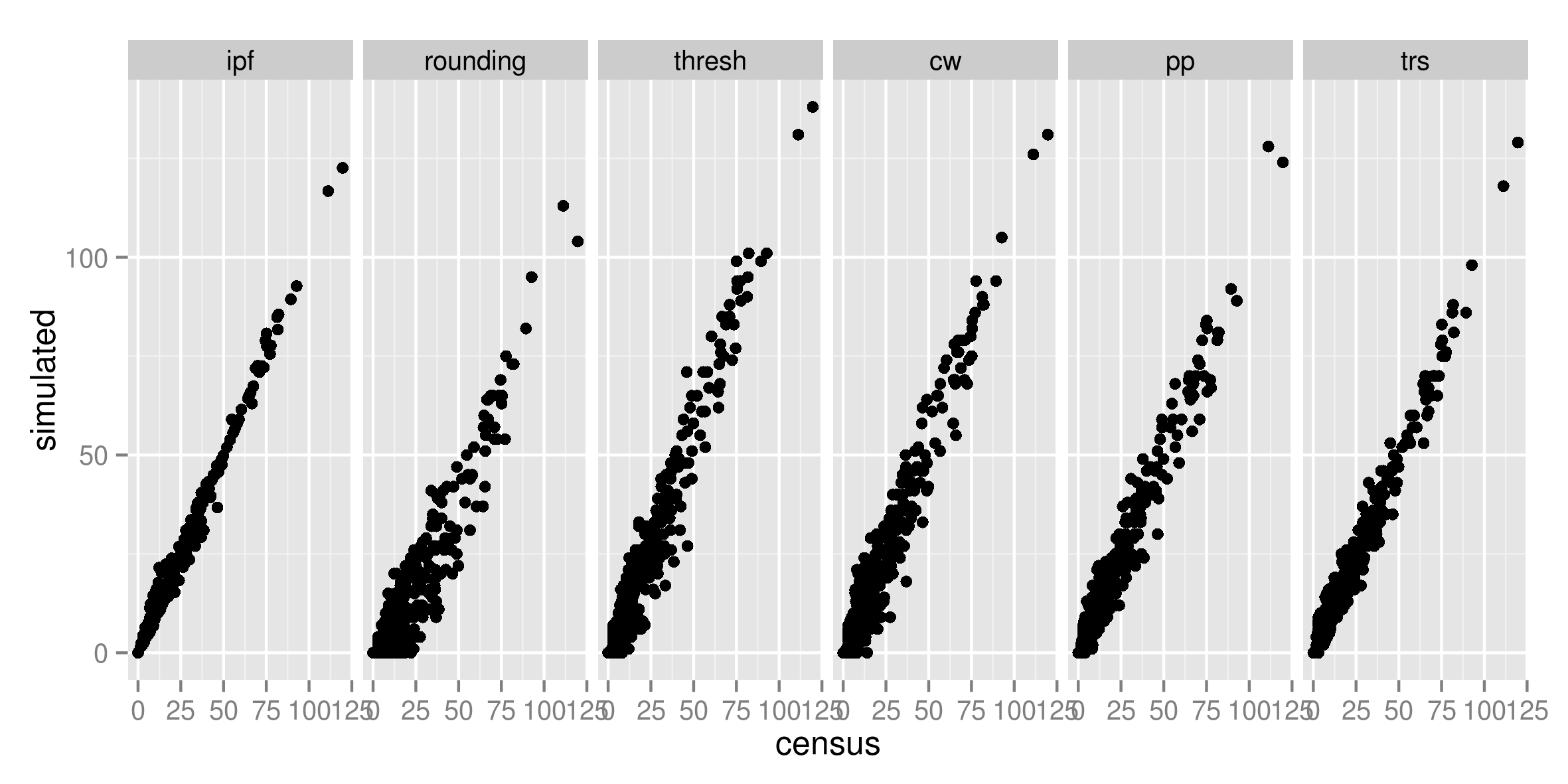}
\caption{Scatter plots illustrating the relationship between
census constraint variables (x axis) and simulated counts for these
variables (y axis) after IPF and four methods of integerising the results.
Each dot represents one variable for one area, 528 dots in each plot
(22 variables multiplied by 24 areas).}
\label{scatter-plots}
\end{figure}

These results can be visualised in scatter plots of census vs simulated
results (Fig.~\ref{scatter-plots}). This figure can be replicated
using the last section of code in `basic-analysis.R',
provided the packages `reshape2' and `ggplot2' have been installed.

We encourage users to test the integerisation methods described
in this user manual on a wider range of
datasets, citing the authors where appropriate.
This will help to check the replicability of the results
presented in the paper that accompanies this code.
It is also hoped that the code and the findings
will be of use to researchers developing, evaluating
and using spatial microsimulation models.

Any feedback would be gratefully
received by robin.lovelace at shef.ac.uk. There is also the
possibility to clone, branch and commit to
a larger code development project related to this research:
\url{https://github.com/Robinlovelace/IPF-performance-testing}.

\section{Reference}
Lovelace, R., \& Ballas, D. (n.d.).  ``Truncate , replicate , sample'': a method
for creating integer weights for spatial microsimulation. Computers,
Environment and Urban Systems. (In press).